\documentclass[twocolumn,numberedappendix, twocolappendix, apj]{openjournal}

\usepackage[english,english]{babel}
\usepackage{amsmath}
\usepackage[usenames]{color}
\usepackage{soul}
\usepackage{graphicx}
\usepackage{float}
\usepackage{bm}
\usepackage{comment}
\usepackage{xcolor}
\usepackage{enumitem}
\usepackage{physics}

\usepackage{longtable}
\usepackage{fontawesome5}

\usepackage{orcidlink}

\usepackage{hyperref}
\hypersetup{colorlinks=true, linkcolor=blue, citecolor=blue, filecolor=black, urlcolor=blue}

\newcommand\blfootnote[1]{
    \begingroup
    \renewcommand\thefootnote{}\footnote{#1}
    \addtocounter{footnote}{-1}
    \endgroup
}

\begin{document}

\title{Spectroscopic Analysis of Pictor II: a very low metallicity ultra-faint dwarf galaxy bound to the Large Magellanic Cloud$^\star$}

\shorttitle{Pictor~II: Stellar Kinematics and Chemical Abundances}


\author{A.~B.~Pace\,\orcidlink{0000-0002-6021-8760}$^{1,\dagger,\ddagger}$ } 
\author{T.~S.~Li\,\orcidlink{0000-0002-9110-6163}$^{2,3}$ }
\author{A.~P.~Ji\,\orcidlink{0000-0002-4863-8842}$^{4,5,6}$ }
\author{J.~D.~Simon\,\orcidlink{0000-0002-4733-4994}$^{7}$ }
\author{W.~Cerny\,\orcidlink{0000-0003-1697-7062}$^{8}$ }
\author{A.~M.~Senkevich\,\orcidlink{0009-0004-5519-0929}$^{9}$ }
\author{A.~Drlica-Wagner\,\orcidlink{0000-0001-8251-933X}$^{10,5,4,6}$ }
\author{K.~Bechtol\,\orcidlink{0000-0001-8156-0429}$^{11}$ }
\author{C.~Y.~Tan\,\orcidlink{0000-0003-0478-0473}$^{5,12}$ }
\author{A.~Chiti\,\orcidlink{0000-0002-7155-679X}$^{4,5}$ }
\author{D.~Erkal\,\orcidlink{0000-0002-8448-5505 }$^{9}$ }
\author{C.~E.~Mart\'inez-V\'azquez\,\orcidlink{0000-0002-9144-7726}$^{13}$ }
\author{P.~S.~Ferguson\,\orcidlink{0000-0001-6957-1627}$^{14}$ }
\author{R.~G.~Kron\,\orcidlink{0000-0003-2643-7924}$^{4}$ }

\author{K~R.~Atzberger\,\orcidlink{0000-0001-9649-8103}$^{1}$ }
\author{A.~Chaturvedi\,\orcidlink{0000-0001-5143-1255}$^{9}$ }
\author{J.~A.~Frieman\,\orcidlink{0000-0003-4079-3263}$^{4,5,15}$ }
\author{N.~Kallivayalil\,\orcidlink{0000-0002-3204-1742}$^{1}$ }
\author{G.~Limberg\,\orcidlink{0000-0002-3204-1742}$^{5}$ }

\author{G.~E.~Medina\,\orcidlink{0000-0002-9269-8287}$^{2,3}$ }
\author{V.~M.~Placco\,\orcidlink{0000-0003-4479-1265}$^{16}$ }
\author{A.~H.~Riley\,\orcidlink{0000-0001-5805-5766}$^{17}$ }
\author{D.~J.~Sand\,\orcidlink{0000-0003-4102-380X}$^{18}$ }
\author{G.~S.~Stringfellow\,\orcidlink{0000-0003-1479-3059}$^{19}$}
\author{R.~P.~van~der~Marel\,\orcidlink{0000-0001-7827-7825}$^{20,21}$ }
\author{J.~A.~Carballo-Bello\,\orcidlink{0000-0002-3690-105X}$^{22}$ }
\author{Y.~Choi\,\orcidlink{0000-0003-1680-1884}$^{16}$ }
\author{D.~Crnojevi\'c\,\orcidlink{0000-0002-1763-4128}$^{23}$ }
\author{P.~Massana\,\orcidlink{0000-0002-8093-7471}$^{24}$ }
\author{B.~Mutlu-Pakdil\,\orcidlink{0000-0001-9649-4815}$^{25}$ }
\author{M.~Navabi\,\orcidlink{0000-0001-9438-5228}$^{9}$ }
\author{N.~E.~D. No\"el\,\orcidlink{0000-0002-8282-469X}$^{9}$ }
\author{J.~D.~Sakowska\,\orcidlink{0000-0002-1594-1466}$^{26}$ }


\affil{$^{1}$ Department of Astronomy, University of Virginia, 530 McCormick Road, Charlottesville, VA 22904, USA}
\affil{$^{2}$ David A. Dunlap Department of Astronomy \& Astrophysics, University of Toronto, 50 St George Street, Toronto ON M5S 3H4, CA}
\affil{$^{3}$ Dunlap Institute for Astronomy \& Astrophysics, University of Toronto, 50 St George Street, Toronto, ON M5S 3H4, CA}
\affil{$^{4}$ Department of Astronomy \& Astrophysics, University of Chicago, Chicago, IL 60637, USA}
\affil{$^{5}$ Kavli Institute for Cosmological Physics, University of Chicago, Chicago, IL 60637, USA}
\affil{$^{6}$ NSF-Simons AI Institute for the Sky (SkAI), 172 E. Chestnut St., Chicago, IL 60611, USA}
\affil{$^{7}$ Observatories of the Carnegie Institution for Science, 813 Santa Barbara St., Pasadena, CA 91101, USA}
\affil{$^{8}$ Department of Astronomy, Yale University, New Haven, CT 06520, USA}
\affil{$^{9}$ School of Mathematics and Physics, University of Surrey, Guildford, Surrey, GU2 7XH, UK}
\affil{$^{10}$ Fermi National Accelerator Laboratory, P.O.\ Box 500, Batavia, IL 60510, USA}
\affil{$^{11}$ Physics Department, 2320 Chamberlin Hall, University of Wisconsin-Madison, 1150 University Avenue Madison, WI  53706-1390}
\affil{$^{12}$ Department of Physics, University of Chicago, Chicago, IL 60637, USA}
\affil{$^{13}$ NSF NOIRLab, 670 N. A'ohoku Place, Hilo, Hawai'i, 96720, USA}
\affil{$^{14}$ DiRAC Institute, Department of Astronomy, University of Washington, 3910 15th Ave NE, Seattle, WA, 98195, USA}
\affil{$^{15}$ SLAC National Laboratory, 2575 Sand Hill Rd, Menlo Park, CA 94025}
\affil{$^{16}$ NSF NOIRLab, 950 N. Cherry Ave., Tucson, AZ 85719, USA}
\affil{$^{17}$ Institute for Computational Cosmology, Department of Physics, Durham University, South Road, Durham DH1 3LE, UK}
\affil{$^{18}$ Steward Observatory, University of Arizona, 933 North Cherry Avenue, Tucson, AZ 85721-0065, USA}
\affil{$^{19}$ Center for Astrophysics and Space Astronomy, University of Colorado Boulder, Boulder, CO 80309, USA}
\affil{$^{20}$ Space Telescope Science Institute, 3700 San Martin Drive, Baltimore, MD 21218, USA}
\affil{$^{21}$ Center for Astrophysical Sciences, The William H. Miller III Department of Physics \& Astronomy, Johns Hopkins University, Baltimore, MD 21218, USA}
\affil{$^{22}$ Instituto de Alta Investigaci\'on, Universidad de Tarapac\'a, Casilla 7D, Arica, Chile}
\affil{$^{23}$ Department of Physics and Astronomy, University of Tampa, 401 West Kennedy Boulevard, Tampa, FL 33606, USA}
\affil{$^{24}$ NSF NOIRLab, Casilla 603, La Serena, Chile}
\affil{$^{25}$ Department of Physics and Astronomy, Dartmouth College, Hanover, NH 03755, USA}
\affil{$^{26}$ Instituto de Astrofísica de Andalucía, CSIC, Glorieta de la Astronom\'\i a,  E-18080 Granada, Spain}

\collaboration{DELVE \& MAGIC Collaborations}

\begin{abstract}
We present  Magellan/IMACS and Magellan/MIKE spectroscopy  of the ultra-faint dwarf (UFD) galaxy  Pictor~II (Pic~II) that is located only 12 kpc from the Large Magellanic Cloud (LMC). 
From the IMACS spectroscopy, we identify   13  member stars and  measure a mean heliocentric  velocity of $326.9\pm1.1~{\rm km~s^{-1}}$, a velocity dispersion of $3.5_{-0.9}^{+1.1}~{\rm km~s^{-1}}$, a mean metallicity of  $\overline{{\rm [Fe/H]}}=-2.99\pm0.06$, and an upper limit on the metallicity dispersion of $\sigma_{\rm [Fe/H]}<0.18$.
We measure detailed elemental abundances for the brightest star, finding $\mbox{[Fe/H]} = -3.3$, high [$\alpha$/Fe] ratios, and no detectable neutron capture elements, similar to stars in other UFDs. However, this star has an unusually high [Sc/Fe] ratio.
The   dynamical mass-to-light ratio ($M/L=760_{-420}^{+910}~M_{\odot}~L^{-1}_{\odot}$), size, and chemical abundances  confirms that Pic~II is a dark matter-dominated dwarf galaxy. 
We perform detailed  orbit modeling of Pic~II in a combined Milky Way (MW) and LMC potential and find that Pic~II is highly likely to be a long-term   LMC satellite.
Furthermore, we find that Pic II is likely still bound to the LMC today.
Pic~II is the seventh LMC-associated UFD and among the most metal-poor UFDs known. 
We further update the morphological parameters with deeper Dark Energy Camera (DECam) photometry, compute the dark matter properties for dark matter indirect detection searches, verify the extremely low   metallicity with narrowband  CaHK imaging, and briefly discuss tidal influences of the LMC and MW. 
\end{abstract}

\maketitle

\section{Introduction}
\label{section:intro}

The advent of wide and deep optical surveys has precipitated the discovery of a large number of ultra-faint dwarf galaxies (UFDs, $M_V \gtrsim -7.7$) around the Milky Way (MW) \citep[e.g.,][]{Willman2005AJ....129.2692W, Belokurov2007ApJ...654..897B, Laevens2015ApJ...813...44L, Bechtol2015ApJ...807...50B, Koposov2015ApJ...805..130K, DrlicaWagner2015ApJ...813..109D, Cerny2023ApJ...953....1C, Smith2023AJ....166...76S, Homma2024PASJ...76..733H, Tan2025ApJ...979..176T}.
UFDs are among the oldest, most chemically pristine, and dark matter dominated galaxies known \citep{McConnachie2012AJ....144....4M, Simon2019ARA&A..57..375S, Pace2024arXiv241107424P} and are excellent probes to test $\Lambda$CDM at small scales \citep{Bullock2017ARA&A..55..343B}.
UFDs are old and metal-poor, and their stars trace chemical enrichment from the first Population~III stars \citep{Simon2007ApJ...670..313S, Brown2014ApJ...796...91B, Frebel2015ARA&A..53..631F}.
The population of UFDs  probes galaxy formation before the epoch of reionization \citep[e.g.,][]{Bose2018ApJ...863..123B} and constrains the small-scale power spectrum of dark matter \citep[e.g.,][]{Jethwa2018MNRAS.473.2060J, Newton2018MNRAS.479.2853N, Nadler2021PhRvL.126i1101N}.

\blfootnote{$^\star$This paper includes data gathered with the 6.5 meter Magellan Telescopes located at Las Campanas Observatory, Chile.}
\blfootnote{$^\dagger$\href{mailto:pvpace1@gmail.com}{pvpace1@gmail.com} or \href{mailto:apace@virginia.edu}{apace@virginia.edu}}
\blfootnote{$^{\ddagger}$ Galaxy Evolution and Cosmology (GECO) Fellow}

The MW's most massive companion galaxies, the Large and Small Magellanic Clouds (LMC/SMC), also have had an impact on the UFD population.
This pair of galaxies constitute ${\sim}$10-15\% of the MW's dark halo mass \citep[e.g.,][]{Erkal2019MNRAS.487.2685E, Koposov2023MNRAS.521.4936K}, and 
they are likely on their first infall into the MW \citep{Besla2010ApJ...721L..97B, Kallivayalil2013ApJ...764..161K}.
The LMC and SMC are predicted to have brought  their own group of faint satellite galaxies with them \citep[e.g.,][]{DOnghia2008ApJ...686L..61D, Deason2015MNRAS.453.3568D, Jethwa2016MNRAS.461.2212J}, thus   increasing the total number of MW satellites \citep[e.g.,][]{Dooley2017MNRAS.472.1060D, Newton2018MNRAS.479.2853N, Nadler2020ApJ...893...48N}.
Evidence of this accreted population can be inferred from the clustering of MW satellites around the LMC and SMC in the Dark Energy Survey (DES) footprint \citep{Koposov2015ApJ...805..130K, DrlicaWagner2015ApJ...813..109D}.

The Magellanic SatelLites Survey\footnote{MagLiteS was a precursor survey to the DECam Local Volume Exploration survey.} (MagLiteS; PI: K. Bechtol) was  designed to search for additional satellites near the Magellanic Clouds \citep{DrlicaWagner2016ApJ...833L...5D}, particularly on the southern side of the LMC/SMC that the DES did not cover. 
Three UFD candidates were discovered with MagLiteS: Pictor~II (Pic~II), Carina~II (Car~II), and Carina~III (Car~III) \citep{DrlicaWagner2016ApJ...833L...5D, Torrealba2016MNRAS.459.2370T}.
The radial velocities of Car~II and Car~III \citep{Li2018ApJ...857..145L} suggested they were associated with the LMC based on simulated LMC satellites
\citep{Jethwa2016MNRAS.461.2212J} and this  association was confirmed with full 6-D phase space information from {\it Gaia} DR2 proper motion measurements \citep{GaiaCollaboration2016A&A...595A...1G, GaiaBrown2018A&A...616A...1G} and orbit models that included the LMC \citep{Erkal2020MNRAS.495.2554E, Patel2020ApJ...893..121P}. 
Several other UFDs, Horologium~I (Hor~I), Hydrus~I (Hyi~I), Phoenix~II (Phe~II), and Reticulum~II (Ret~II), have additionally been associated with the LMC \citep{Koposov2018MNRAS.479.5343K, Simon2018ApJ...863...89S, Kallivayalil2018ApJ...867...19K, Erkal2020MNRAS.495.2554E, Patel2020ApJ...893..121P, CorreaMagnus2022MNRAS.511.2610C, Battaglia2022A&A...657A..54B, Pace2022ApJ...940..136P, Vasiliev2023Galax..11...59V}.
While some of these UFDs may be SMC satellites, as the SMC is a satellite of the LMC,  we only consider the LMC connection in this work.

The LMC UFD population enables studying UFDs that formed in a different and lower mass  environment compared to the in-situ MW UFDs. 
Star formation history measurements suggest that on average the  LMC satellites  were quenched $\sim600~{\rm Myr}$ more recently compared to MW satellites \citep{Sacchi2021ApJ...920L..19S, Durbin2025arXiv250518252D}.  This may be due to patchy reionization allowing lower density environments to form stars longer \citep[e.g.,][]{Lunnan2012ApJ...746..109L, Katz2020MNRAS.494.2200K, Kim2023ApJ...959...31K}. 
Conversely, no difference has been seen in the size-luminosity plane between the LMC and MW UFDs \citep{Richstein2024ApJ...967...72R} but the number of LMC confirmed UFDs is an order of magnitude smaller than MW UFDs and confirming more LMC UFDs is key to studying environmental differences.

Pic~II was identified as a highly  probable  LMC-associated system   due to its close proximity to the LMC \citep[$d_{\rm LMC}\sim11~{\rm kpc}$;][]{DrlicaWagner2016ApJ...833L...5D}, but has not yet been studied spectroscopically.
The properties of Pic~II are summarized in Table~\ref{table:properties}.
Here we present the first stellar spectroscopy   of Pic~II to assess its nature and potential association with the LMC.
In Section~\ref{section:data}, we present our observational data including new  Magellan/IMACS (\S~\ref{section:imacs}) and Magellan/MIKE (\S~\ref{section:mike}) observations.
In Section~\ref{section:properties}, we derive the systemic properties of Pic~II including: updated structural and morphological parameters (\S~\ref{section:structure}), stellar kinematics (\S~\ref{section:stellar_kinematics}), and the spectroscopic metallicity distribution (\S~\ref{section:metallicity_dispersion}).
In Section~\ref{section:discussion}, we discuss the nature of Pic~II (\S~\ref{section:nature}), derive the orbit and discuss the association with the LMC (\S~\ref{section:lmc}), 
discuss the detailed chemical abundances of the brightest star (\S~\ref{section:abundances}), and determine the dark matter properties of Pic~II (\S~\ref{section:j_factor}).
We summarize our findings in Section~\ref{section:summary}.

\begin{deluxetable*}{l c c c c }
\tablecolumns{4}
\tablewidth{0pt}
\tabletypesize{\footnotesize}
\tablecaption{\label{table:properties}
Photometric and kinematic properties of Pic~II.
}
\tablehead{
\colhead{Parameter} & \colhead{Description} & \colhead{Pic~II} & \colhead{Units} & \colhead{Section}
}
\startdata
R.A. & R.A. (J2000) & $101.175_{-0.006}^{+0.006}$ & deg & \ref{section:structure} \\
Decl. & Decl. (J2000)& $-59.901_{-0.003}^{+0.003}$ & deg & \ref{section:structure}\\
$a_h$ & Angular semi-major axis length & $2.8_{-0.3}^{+0.4}$ & arcmin & \ref{section:structure}\\
$a_{1/2}$ & Physical semi-major axis length & $36.1_{-4.3}^{+5.7}$  & pc & \ref{section:structure}\\
$R_h$ & Azimuthally-averaged angular half-light radius  & $2.4_{-0.3}^{+0.3}$ & arcmin & \ref{section:structure}\\
$R_{1/2}$ & Azimuthally-averaged physical half-light radius & $32.0_{-3.6}^{+4.4}$ & pc & \ref{section:structure}\\
$\epsilon$ & Ellipticity & $0.22_{-0.13}^{+0.11}$ & & \ref{section:structure}\\
P.A. & Position angle of major axis (defined east of north) & $55_{-18}^{+20}$  & deg & \ref{section:structure}\\
$(m-M)_0$ & Distance modulus & $18.26_{-0.03}^{+0.04} \pm 0.1$ $^a$ &  & \ref{section:structure}\\
$D_{\odot}$ & Heliocentric distance  & $44.9_{-0.6}^{+0.9}\pm 2.0$ $^a$ & kpc & \ref{section:structure}\\
$M_V$ & Absolute (integrated) V-band magnitude & $-2.65_{-0.48}^{+0.69}$  & mag & \ref{section:structure}\\
$L_V$ & V-band luminosity &  $970_{-370}^{+630}$& $L_{\odot}$ & \ref{section:structure}\\
$M_\star$ & Stellar mass (assuming $M_\star/L_V$ = 2) &  $1950_{-750}^{+1260}$ & $M_{\odot}$ & \ref{section:structure}\\
$E(B-V)$ & Mean reddening within the half-light radius & 0.11 & mag & \ref{section:structure}\\
$D_{\rm LMC}$ &  Distance from the LMC & $11.8_{-0.4}^{+0.8}$ & kpc & \ref{section:structure}\\
$D_{\rm GC}$ &   Galactocentric distance & $45.6_{-0.6}^{+0.9}\pm2.0$ $^a$  & kpc & \ref{section:structure}\\

\multicolumn{5}{c}{ \hrulefill } \\

$\overline{v}_{\rm hel}$ & Systemic velocity in the heliocentric frame  & $326.9\pm1.1$ & ${\rm km~s^{-1}}$ & \ref{section:stellar_kinematics} \\
$\overline{v}_{\rm gsr}$ & Systemic velocity in the Galactocentric frame & $104.5\pm 1.1$ & ${\rm km~s^{-1}}$ & \ref{section:stellar_kinematics} \\
$\sigma_v$ & Velocity dispersion &  $3.5_{-0.9}^{+1.1}$  & ${\rm km~s^{-1}}$ & \ref{section:stellar_kinematics} \\
$M_{1/2}$ & Dynamical mass within the half-light radius &  $3.6_{-1.6}^{+2.9} \times 10^5 $ & $M_{\odot}$ & \ref{section:stellar_kinematics}\\
$M/L$ & Dynamical mass-to-light ratio within the half-light radius  & $760_{-420}^{+910}$  & $M_{\odot}~L_{\odot}^{-1}$ & \ref{section:stellar_kinematics}\\
$\overline{{\rm [Fe/H]}}$ & Mean metallicity from spectroscopic CaT & $-2.99 \pm 0.06$  & & \ref{section:metallicity_dispersion}\\
$\sigma_{\rm [Fe/H]}$ & Metallicity dispersion from spectroscopic CaT & $<0.18$ $^b$ &  & \ref{section:metallicity_dispersion}\\
$\overline{\mu}_{\alpha\star}$  & Systemic proper motion in the R.A. direction & $1.18^{+0.05}_{-0.05} \pm 0.023$  $^c$ & ${\rm mas~yr^{-1}}$  & \ref{section:proper_motion}\\
$\overline{\mu}_{\delta}$ & Systemic proper motion in the Decl. direction & $1.16^{+0.05}_{-0.05} \pm0.023$ $^c$  & ${\rm mas~yr^{-1}}$ & \ref{section:proper_motion} \\
$C_{\mu_{\alpha\star}\times\mu_\delta}$ & Correlation between $\mu_{\alpha\star}$ and $\mu_{\delta}$ & -0.035 & & \ref{section:proper_motion}\\ 

$r_{\rm peri}$ & Orbital pericenter relative to MW & $38.9^{+2.5}_{-3.0}$ & kpc & \ref{section:lmc}\\
$r_{\rm apo}$ & Orbital apocenter relative to MW & $219^{+99}_{-54}$ & kpc & \ref{section:lmc}\\ 
$e$ & Orbital eccentricity relative to MW & $0.70^{+0.09}_{-0.08}$ &  & \ref{section:lmc}\\

$r_{\rm peri, \, LMC}$ & Orbital pericenter relative to the LMC & $7.8^{+4.4}_{-3.4}$ & kpc & \ref{section:lmc}\\
$r_{\rm apo, \, LMC}$ & Orbital apocenter relative to the LMC & $29.9^{+9.7}_{-6.0}$ & kpc & \ref{section:lmc}\\ 
$e_{\rm LMC}$ & Orbital eccentricity relative to the LMC & $0.62^{+0.16}_{-0.16}$ &  & \ref{section:lmc}\\
$p_{\rm LMC}$ & Fraction of orbits bound to the LMC & $0.91$ &  & \ref{section:lmc}\\

$\log_{10}{J(0\fdg2)}$ & Integrated J-factor within a solid angle of $0\fdg2$ & $18.33\pm0.53$ & $\log_{10}{\rm GeV^2~cm^{-5}}$ & \ref{section:j_factor} \\
$\log_{10}{J(0\fdg5)}$ & Integrated J-factor within a solid angle of $0\fdg5$  & $18.48\pm0.55$ & $\log_{10}{\rm GeV^2~cm^{-5}}$ & \ref{section:j_factor}  
\enddata
\tablecomments{}
\tablenotetext{a}{A systematic error of 0.1 is included in the distance modulus which corresponds to a systematic error of $\sim 2~{\rm kpc}$  in the heliocentric and Galactocentric distance.}
\tablenotetext{b}{Upper limits listed here are at the 95\% credible level.}
\tablenotetext{c}{The last term corresponds to the systematic proper motion error  from \citet{Gaia_Lindegren_2021A&A...649A...2L}.}

\end{deluxetable*}

\section{Data}
\label{section:data}

\subsection{Medium-resolution spectroscopy: Stellar Kinematics and Metallicities}
\label{section:imacs}

\subsubsection{Target Selection and Observations}

\begin{deluxetable*}{@{}lllllcrcccc@{}}
\tablecaption{Summary of IMACS observations.\label{table:observing}}
\tablehead{
\colhead{MJD$^{a}$} & \colhead{Run} & \colhead{Mask}   &   \colhead{R.A. (h:m:s)}    & \colhead{Decl. (d:m:s)}   &\colhead{$\sum{t_{\rm exp}}~({\rm s})$} & \colhead{Seeing} & \colhead{\# targets} & \colhead{\# useful$^{b}$}  }
\startdata
57779.17 & Jan 2017 & Pic2\_mask1 & 06:44:50.000 & -59:54:15.00     &  14400           & 1\farcs5    & 45          & 37   \\
58103.53 & Dec 2017 & Pic2\_mask2 & 06:44:45.100 & -59:55:22.00   & 34200   & 0\farcs65-1\farcs25     & 40         & 27   \\ 
58104.13 &  Dec 2017 & Pic2\_mask3 & 06:44:50.000 & -59:53:35.00   & 9600   & 0\farcs75          & 26          & 19   \\
58499.11 & Jan 2019 & Pic2\_mask4 & 06:44:32.000 & -59:57:00.00   & 9000    & 0\farcs6-1\farcs1         & 19          & 19   \\ 
58526.54 & Feb 2019 & Pic2\_mask5 & 06:44:13.800 & -59:47:00.00   & 7346  & 0\farcs65-2\farcs0         & 13          & 10   
\enddata  
\tablenotetext{a}{The MJD presented is the average over multiple nights of a single run (where applicable).}
\tablenotetext{b}{The number of observed spectra with a reliable radial velocity measurement.}
\end{deluxetable*}

Pic~II was observed using Inamori Magellan Areal Camera and Spectrograph (IMACS) \citep{Dressler2006SPIE.6269E..0FD, Dressler2011PASP..123..288D}, a multi-object slit spectrograph at the Magellan Baade 6.5m telescope over 4 observing runs spanning Jan 2017 to Feb 2019. 
Our observations are summarized in Table~\ref{table:observing}.
We used the same instrument configuration as our previous IMACS observations of UFD galaxy candidates \citep{Simon2017ApJ...838...11S, Li2017ApJ...838....8L, Li2018ApJ...857..145L, Simon2020ApJ...892..137S, Bruce2023ApJ...950..167B, Cerny2023ApJ...942..111C}. 
Specifically, we observed the spectral region 7500--9000~\AA~at resolution R $\sim$ 11,000 and an effective field-of-view of $8\arcmin \times 15\arcmin$ with the f/4 camera, the 1200/32\fdg7 grating, and a slit width of $0\farcs7$.

Target selection was performed using MagLiteS photometry~\citep{DrlicaWagner2016ApJ...833L...5D}. We selected potential members of Pic~II as those that lie spatially near the center of Pic~II and photometrically near a PARSEC isochrone \citep{Bressan2012MNRAS.427..127B} with [Fe/H] = $-2.2$ and age = 12.5 Gyr at distance modulus of $m-M = 18.3$ as measured by~\citet{DrlicaWagner2016ApJ...833L...5D}. 
In 2019, we added {\it Gaia} astrometry to our target selection and included  high probability members using the proper motion information provided by {\it Gaia} DR2 \citep{GaiaBrown2018A&A...616A...1G}, using the methods described in \citet{Pace2019ApJ...875...77P}.

The IMACS data were reduced as described by \citet{Simon2017ApJ...838...11S} and \citet{Li2017ApJ...838....8L, Li2018ApJ...866...22L}, employing a combination of the Cosmos reduction pipeline \citep{Dressler2011PASP..123..288D, Oemler2017ascl.soft05001O} and a version of the DEEP2 data reduction pipeline \citep{Cooper2012ascl.soft03003C, Newman2013ApJS..208....5N} adapted for IMACS.
In order to keep the temporal information to check binarity as well as to check potential systematics, we only coadded spectra taken with the same mask and within a given $<3$ night run.

\subsubsection{Velocity and Metallicity Measurements}\label{s:measure}

The line-of-sight velocity and metallicity measurements follow the same method as described in other papers by our group \citep[e.g.,][]{Li2017ApJ...838....8L, Li2018ApJ...857..145L, Li2018ApJ...866...22L}. Specifically, the velocities are derived via template fitting with a Markov chain Monte Carlo sampler; and the metallicities are derived by measuring the equivalent widths (EWs) of the Calcium Triplet (CaT) lines around 8500~\AA~ using the calibration from \citet{Carrera2013MNRAS.434.1681C}.
Spectra with signal-to-noise  $>5$  (S/N $>$9) usually have good velocity  (metallicity) measurements. 
Successful velocities and metallicities  are presented in Table~\ref{table:imacs_members} and  per-mask  measurements are included in the zenodo repository\footnote{\url{https://doi.org/10.5281/zenodo.15706700}}.
In total, we  observed 143 targets with 112 good quality velocity measurements of 94 unique stars. We obtained 99 good quality  CaT EW measurements of 84 unique stars.

For our primary analysis, we combined repeat measurements at the catalog level.
For the combined velocities we fit a free velocity and dispersion to the repeat measurements and use the  mean velocity and error for the star’s intrinsic properties. The velocity errors from this method will match the variance of the weighted mean for non-variable stars (the expected behavior for non-variable sources with errors that are Normally distributed) and will be larger for stars with evidence of variability. 
This combination method can be more robust to unidentified binary stars as stars with variation will have larger errors \citep{Buttry2022MNRAS.514.1706B}. 
In principle, the center-of-mass velocity for binary stars should be used but this is extremely difficult to correctly identify with only a few velocity measurements.
For repeat metallicity measurements we combine measurements with the weighted mean and variance of the weighted mean for the combined error.
We have verified that using the best measurement (highest signal-to-noise) instead of the combined value gives similar results (when excluding binary stars).

\begin{deluxetable*}{cccccccccc}

\tablecaption{Properties of spectroscopically confirmed member stars  of Pic~II\label{table:imacs_members}}
\tablehead{ID & Gaia DR3 source\_id & RA & DEC & $g_0$ & $r_0$ &  $v_{\rm hel}$ &  $\Sigma$ EW CaT & [Fe/H]   \\
    & & (deg) & (deg) & (mag)&  (mag) &  (km s$^{-1}$)  & (\r{A})  & (dex)  }
\startdata
10286300071032 & 5480249356255194112 & 101.22791 & -59.91751 & 17.48 & 16.75 & 329.4$\pm$0.6 & 1.92$\pm$0.12& -3.08$\pm$0.08 \\
10286300171312 & 5480252100736880640 & 101.07264 & -59.92858 & 19.03 & 18.51 & 327.1$\pm$0.6 & 1.46$\pm$0.11& -3.06$\pm$0.09 \\
10286300169004 & 5480170908676338432 & 100.89608 & -59.88540 & 19.19 & 18.71 & 320.5$\pm$1.6 & 1.61$\pm$0.60& -2.91$\pm$0.41 \\
10286300346714 & 5480248011928003968 & 101.28227 & -59.97635 & 18.91 & 19.05 & 330.8$\pm$1.4 & & \\
10286300169363 & 5480252444334270080 & 101.20003 & -59.89651 & 18.86 & 19.08 & 331.0$\pm$1.9 & & \\
10286300169338 & 5480252276833093248 & 101.15539 & -59.89063 & 19.76 & 19.28 & 327.4$\pm$0.9 & 1.41$\pm$0.17& -2.94$\pm$0.14 \\
10286300001346 & 5480252066377147008 & 101.17612 & -59.90296 & 20.08 & 19.63 & 325.3$\pm$1.1 & 1.53$\pm$0.24& -2.79$\pm$0.19 \\
10286300417761 & 5480158092493302272 & 100.94904 & -59.97386 & 20.38 & 19.92 & 328.7$\pm$1.5 & 1.21$\pm$0.28& -3.00$\pm$0.28 \\
10286300169343 & 5480252341255057536 & 101.15827 & -59.88901 & 20.41 & 19.97 & 326.9$\pm$1.7 & 1.42$\pm$0.44& -2.82$\pm$0.36 \\
10286300001343 & 5480252276831909376 & 101.17167 & -59.89087 & 20.70 & 20.25 & 329.1$\pm$7.5 & 1.35$\pm$0.20& -2.74$\pm$0.18 \\
10286300071017 & 5480247840129310848 & 101.18892 & -59.98813 & 20.89 & 20.45 & 317.0$\pm$2.4 & & \\
10286300300125 & 5480249008361992448 & 101.15214 & -59.94265 & 20.96 & 20.53 & 330.9$\pm$2.7 & & \\
10286300171369 &  & 101.10710 & -59.94019 & 21.51 & 21.13 & 326.0$\pm$2.3 & 1.02$\pm$0.81& -2.81$\pm$0.73 
\enddata
\tablecomments{Pic~II spectroscopic members ordered by increasing r-band magnitude. Details of each parameter can be found in Section \ref{section:imacs}. The full spectroscopic sample along with additional columns (e.g., select {\it Gaia} DR3 astrometry, MAGIC/CaHK photometry and metallicity, and additional DELVE photometry) can be found at the following zenodo repository: \url{https://doi.org/10.5281/zenodo.15706700}}
\end{deluxetable*}

\subsection{High-resolution spectroscopy: Chemical Abundance Analysis}\label{section:mike}

\subsubsection{Observations and Radial Velocity}
We obtained Magellan Inamori Kyocera Echelle (MIKE) spectra \citep{Bernstein2003SPIE.4841.1694B} at the Magellan Clay 6.5 m telescope of  star 10286300071032\footnote{Gaia DR3 source\_id = 5480249356255194112}, hereafter PicII-1,
the brightest known member of Pic~II ($G \sim 17.05$), on 2017 December 06.
We used the 1\farcs0 slit which has a spectral resolution of $R \sim 28,000$ and 22,000 in the blue and red arms, respectively.
We obtained 3 x 2700s exposures with 2x2 binning, resulting in S/N per pixel ${\sim}14$ at 4250{\AA} and ${\sim}35$ at 6000{\AA}.
Data were reduced using the standard MIKE pipeline in CarPy \citep{Kelson2003PASP..115..688K}\footnote{\url{http://code.obs.carnegiescience.edu/mike}}.
We analyzed the spectrum using the 2017 version of the 1D LTE radiative transfer code MOOG \citep{Sneden1973PhDT.......180S} updated to include  opacity from scattering \citep{Sobeck2011AJ....141..175S, Sobeck2023ascl.soft08001S}\footnote{\url{https://github.com/alexji/moog17scat}}.
We used the alpha-enhanced plane-parallel \citet{Castelli2003IAUS..210P.A20C} (ATLAS) model atmospheres.
Normalization, radial velocity correction, equivalent width measurements, and spectral synthesis fitting was done with \texttt{smhr}\footnote{\url{https://github.com/andycasey/smhr}}, first described in \citet{Casey2014PhDT.......394C}.

The radial velocity was determined following \citet{Ji2020ApJ...889...27J}.
Each MIKE spectral order from 3700{\AA}-6800{\AA} was cross-correlated against a rest frame spectrum of the metal-poor red giant HD122563. As velocity uncertainties are dominated by wavelength calibration issues, we adopt the order-to-order scatter as the velocity uncertainty. The final velocity is $v_{\rm hel} = 331.0 \pm 1.0$ km s$^{-1}$.
Separately, our IMACS data found PicII-1 to have velocities 330.0, 328.2, 329.9 km s$^{-1}$, each with uncertainty 1 km s$^{-1}$. There is no evidence for large radial velocity variations for PicII-1.

\subsubsection{Stellar Parameters and Abundances}
\label{section:mike_abundances}

Stellar parameters were determined following the procedure in \citet{Frebel2013ApJ...769...57F}, i.e. excitation, ionization, and line strength balance of Fe lines with an empirical correction to match a photometric temperature scale. We find $T_{\rm eff}=4520 \pm 174$\,K, $\log{g} = 0.80 \pm 0.37$, $v_t=2.07 \pm 0.22~{\rm km~s^{-1}}$, and $\mbox{[Fe/H]}=-3.27 \pm 0.22$ for the effective temperature, surface gravity, microturbulence velocity, and metallicity, respectively.
The uncertainties include statistical error from the Fe lines as well as systematic uncertainties of 150\,K, 0.3 dex, and 0.2$~{\rm km~s^{-1}}$ for $T_{\rm eff}$, $\log{g}$, and $v_t$ respectively that reflect scatter in the \citet{Frebel2013ApJ...769...57F} calibration.
Line lists were adopted from \citet{Ji2020ApJ...889...27J}.
Abundances of most elements were determined from equivalent widths, but for the following features we determined the abundance with spectral synthesis: molecular CH bands; hyperfine structure in Sc, Mn, Co; and Al and Si lines somewhat blended with CH.
We only use lines with $\lambda > 4000${\AA} because the S/N is too low at bluer wavelengths ($<8$ per pixel).
The exception is a strong Al line at 3944{\AA} as no other Al lines are available: this line is clearly detected, but given the S/N the abundance is extremely uncertain (constraint no better than 0.5 dex).
We also determined formal $5\sigma$ upper limits for Sr, Ba, and Eu using spectral synthesis of the 4215{\AA}, 4934{\AA}, and 4129{\AA} lines, respectively (see \citealt{Ji2020AJ....160..181J} for more details).

Final abundances and uncertainties of PicII-1 are given in Table~\ref{tbl:abunds}.
For each element, $N$ is the number of lines measured.
$\log \epsilon(X)$ is the average abundance of those lines weighted by the abundance uncertainty. If $\log\epsilon_i$ and $\sigma_i$ are the abundance and uncertainty of line~$i$, then we define $w_i = 1/\sigma_i^2$ and $\log\epsilon(X) = \sum_i (w_i \log\epsilon_i) / \sum_i w_i$.
The uncertainty $\sigma_i$ for an individual line is found by propagating spectrum noise uncertainty.
$\sigma$ is the standard deviation of those lines (undefined  if one line).
$\sigma_{\rm w}$ accounts for the abundance uncertainty due to propagating individual line uncertainties, i.e. $1/\sigma_{\rm w}^2 = \sum_i w_i$ \citep{McWilliam1995AJ....109.2757M}.
$\mbox{[X/H]}$ is the abundance relative to solar abundances from \citet{Asplund2009ARA&A..47..481A}.
$\mbox{[X/Fe]}$ is calculated using either [Fe\,I/H] or [Fe\,II/H], depending on whether X is neutral or ionized.
$\sigma_{\rm XH}$ is the quadrature sum of $\sigma/\sqrt{N}$, $\sigma_{\rm w}$, and abundance uncertainties due to $1\sigma$ stellar parameter variations.
$\sigma_{\rm XFe}$ is similar to $\sigma_{\rm XH}$, but when calculating the stellar parameter uncertainties we include variations in Fe.
We use the difference in Fe\,I abundance for neutral species and the difference in Fe\,II abundance for ionized species to calculate this error.
The [X/Fe] error is usually smaller than the [X/H] error, since abundance differences from changing $T_{\rm eff}$ and $\log{g}$ usually affect X and Fe in the same direction.
All analyses are conducted in LTE, but non-LTE effects are significant for many elements \citep[e.g.,][]{Sitnova2021MNRAS.504.1183S}. Due to inhomogeneities in how non-LTE corrections are determined and applied, we reserve a full discussion of this for a future analysis of all UFD stars. However, the most significant effects in cool giants are that Al and Mn increase by 0.5-1.0 dex \citep{Nordlander2017A&A...607A..75N,Bergemann2019}; Na and K decrease by up to 0.5 dex \citep{Lind2011A&A...528A.103L,Reggiani2019}. Other elements change by $<0.2$ dex \citep{Sitnova2021MNRAS.504.1183S}.

\begin{deluxetable*}{lrrrrrrrr}
\tablecolumns{9}
\tablewidth{0pt}
\tabletypesize{\footnotesize}
\tablecaption{Abundances of PicII-1 \label{tbl:abunds}}
\tablehead{\colhead{El.} & \colhead{$N$} & \colhead{$\log \epsilon(X)$} & \colhead{$\sigma_{\rm w}$} & \colhead{$\sigma$} & \colhead{$\mbox{[X/H]}$} & \colhead{$\sigma_{\rm XH}$} & \colhead{$\mbox{[X/Fe]}$} & \colhead{$\sigma_{\mbox{XFe}}$}}
\startdata
CH    &   2 &  4.80 & 0.04  & 0.04  & -3.63 & 0.48  & -0.36 & 0.21 \\
Na I  &   1 &  3.96 & 0.06  &\nodata& -2.28 & 0.27  &  0.99 & 0.12 \\
Mg I  &   5 &  5.24 & 0.07  & 0.15  & -2.36 & 0.23  &  0.91 & 0.10 \\
Al I  &   1 &  2.82 &\nodata&\nodata& -3.63 &\nodata& -0.36 &\nodata\\
Si I  &   2 &  4.87 & 0.20  & 0.12  & -2.64 & 0.44  &  0.64 & 0.21 \\
K I   &   2 &  2.89 & 0.04  & 0.03  & -2.14 & 0.17  &  1.13 & 0.07 \\
Ca I  &   8 &  3.47 & 0.06  & 0.15  & -2.87 & 0.15  &  0.40 & 0.10 \\
Sc II &   7 &  0.46 & 0.09  & 0.09  & -2.69 & 0.18  &  0.59 & 0.18 \\
Ti II &  18 &  2.15 & 0.04  & 0.14  & -2.80 & 0.15  &  0.48 & 0.08 \\
Cr I  &   5 &  2.15 & 0.09  & 0.16  & -3.49 & 0.25  & -0.22 & 0.10 \\
Mn I  &   2 &  1.04 & 0.18  & 0.18  & -4.39 & 0.34  & -1.12 & 0.19 \\
Fe I  &  96 &  4.23 & 0.02  & 0.22  & -3.27 & 0.21  &\nodata&\nodata\\
Fe II &  11 &  4.22 & 0.07  & 0.21  & -3.28 & 0.14  &\nodata&\nodata\\
Co I  &   2 &  1.81 & 0.11  & 0.11  & -3.18 & 0.29  &  0.09 & 0.13 \\
Ni I  &   1 &  2.71 & 0.11  &\nodata& -3.51 & 0.24  & -0.24 & 0.12 \\
Sr II &   1 &$<-2.06$&\nodata&\nodata&$<-4.93$&\nodata&$<-1.65$&\nodata\\
Ba II &   1 &$<-2.68$&\nodata&\nodata&$<-4.86$&\nodata&$<-1.58$&\nodata\\
Eu II &   1 &$<-2.05$&\nodata&\nodata&$<-2.57$&\nodata& $<0.71$&\nodata
\enddata
\tablecomments{ID=10286300071032 and Gaia DR3 source\_id = 5480249356255194112. The Al abundance has an uncertainty $>0.5$ dex. The CH abundance has not been corrected for the star's evolutionary status here, but the correction is $+0.73$\,dex \citep{Placco2014ApJ...797...21P}.}
\end{deluxetable*}

\subsection{DELVE photometry}

We  utilize data from the  DECam Local Volume Exploration Survey  \citep[DELVE;][]{DrlicaWagner2021ApJS..256....2D},  a southern sky  survey performed with  DECam \citep{Flaugher2015AJ....150..150F} on the 4-m NSF Víctor M. Blanco Telescope at the NSF Cerro Tololo Inter-American Observatory (CTIO).  For this analysis, we used the  stellar catalog from the upcoming   Third Data Release (DR3) of DELVE (Drlica-Wagner, DELVE Collaboration in prep.), which utilizes the DES Data Management  data processing pipeline \citep{Morganson2018PASP..130g4501M} as described by \citet{Tan2025ApJ...979..176T} and \citet{Anbajagane2025arXiv250217674A}. The data release covers $\sim$14,000 deg$^2$ of the sky with a median limiting magnitude of $g \sim$ 24.1, $r \sim$ 23.6, $i \sim$ 23.2, $z \sim$ 22.5 (estimated at S/N = 10 in 2" aperture) and the DELVE catalog contains the MagLiteS observations used to discover Pic~II. 
The DECam data were processed through the image coaddition pipeline closely following the procedure described for DES DR2 \citep{DES2021ApJS..255...20A}. The survey uses \texttt{SourceExtractor} to perform object detection on the $r$ + $i$ + $z$ detection coadd image, with a detection threshold of $\sim5\sigma$. We use photometry derived using multi-epoch PSF model fits which are obtained using \texttt{fitvd} \citep{Hartley2022MNRAS.509.3547H}.   \texttt{fitvd} obtained fluxes for each detected source by fitting  individual-epoch PSF models at the individual constituent images that went into the coadd at the object's location. 
For star-galaxy separation we use $0 \leq \texttt{EXT\_FITVD} \leq 2$ \citep{Bechtol2025arXiv250105739B}.
We note the photometry presented in this paper is corrected for extinction using the $E(B-V)$ values from the reddening map of \citet{Schlegel1998ApJ...500..525S} with the correction from \citet{Schlafly2011ApJ...737..103S}.
We find the  mean reddening within the half-light radius of Pic~II to be $E(B-V) = 0.11$.

\subsection{MAGIC Metallicity Sensitive Narrowband Photometry}

Pic~II was imaged with the narrow-band N395 filter on DECam as part of the Mapping the Ancient Galaxy in CaHK Survey (DECam MAGIC; Chiti et al., in prep) with NOIRLab Prop. ID 2023B-646233. 
The N395 filter includes the Ca II H and K lines at 3968.5~{\AA} and 3933.7~{\AA}, respectively, similar to other metallicity-sensitive filters \citep[e.g., the Pristine survey;][]{Starkenburg2017MNRAS.471.2587S}, allowing the derivation of reasonably precise metallicities \citep[$\sigma_{\rm [Fe/H]}\sim ~0.16$;][]{Barbosa2025arXiv250403593B} for individual stars through photometry.
Specifically, MAGIC obtained 3$\times$12\,min exposures that were dithered and centered on Pic~II on the night of February 7th, 2024\footnote{Our analysis uses the internal MAGIC version v250130}. 
More details on the processing and photometric metallicities will be presented in Chiti et al., (in prep), but we refer the reader to \citet{Barbosa2025arXiv250403593B} for an overview and early result from the MAGIC survey. In particular, see Figure 4 of \citeauthor{Barbosa2025arXiv250403593B} for metallicity error calibration.
Specific to this work, we use this narrow-band CaHK data to verify the metal-poor nature of Pic~II members as an additional check on their membership.
Lastly we note that the MAGIC catalog uses DELVE DR2 photometry for the metallicity derivation and future MAGIC catalogs will utilize the upcoming DELVE DR3 \citep{DELVE2022ApJS..261...38D}.

\subsection{Gaia Astrometry}

We use  {\it Gaia} DR3 astrometry \citep{Gaia_Lindegren_2021A&A...649A...2L, Gaia_Vallenari2023A&A...674A...1G} to assist in Pic~II membership and to measure the systemic proper motion of Pic~II.
We apply the following cuts to select stars with good quality astrometry: \texttt{astrometric\_params\_solved}$>3$, \texttt{ruwe}$<1.4$, and \texttt{astrometric\_excess\_noise\_sig} $< 2$ \citep[following][]{Gaia_Lindegren_2021A&A...649A...2L, Pace2022ApJ...940..136P}.

\section{Systemic Properties of Pic~II}
\label{section:properties}

\begin{figure*}
\centering
\includegraphics[width=0.95\textwidth]{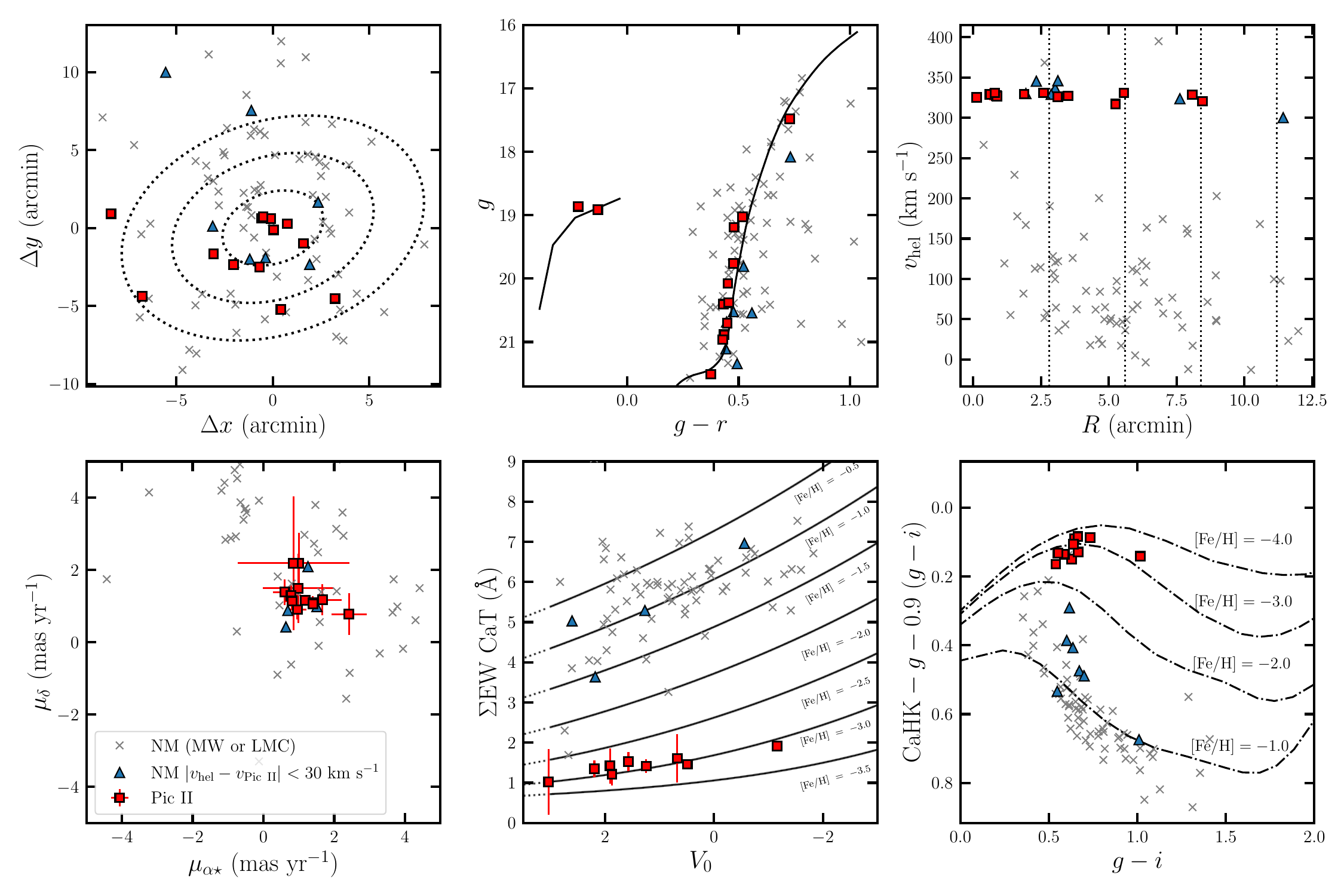}
\caption{
Chemo-dynamic properties of the Pic~II spectroscopic sample. Pic~II members are shown in red squares while non-members with a velocity near Pic~II are in blue triangles ($|v_{\rm hel}-v_{\rm Pic~II}|<30~{\rm km~s^{-1}}$) and the remainder of the non-members are shown in gray x's (NM in legend). {\bf Upper-left}: spatial position of the sample relative to the center of Pic~II.  Dotted ellipses correspond to 1,2,3  $\times R_{h}$. 
{\bf Upper-middle}: DELVE DR2 color-magnitude diagram ($g-r$ vs $g$). We include an old, metal-poor red-giant branch isochrone (age = 12.5~Gyr; Z=0.0001) from \citet{Dotter2008ApJS..178...89D} and the  blue-horizontal branch track  of the old, metal-poor globular cluster M92 scaled to the distance of Pic~II.  {\bf Upper-right}: projected radius ($R$) vs heliocentric velocity ($v_{\rm hel}$). Dotted vertical lines are multiples of the half-light radius.
{\bf Lower-left}: Vector point diagram ($\mu_{\alpha \star}$ vs $\mu_\delta$). {\bf Lower-middle}: V-band absolute magnitude vs Calcium Triplet (CaT) equivalent width. Lines of constant [Fe/H] from \citet{Carrera2013MNRAS.434.1681C} are overlaid. Note that the V-band absolute magnitude assumes the Pic~II distance modulus and only Pic~II members will  have the correct absolute magnitude. Regardless, there is a clear difference in the equivalent width between Pic~II members and MW/LMC foreground stars.
{\bf Lower-right}: metallicity-sensitive color-color diagram with MAGIC CaHK photometry. Lines of constant [Fe/H]  with $\log{g}=2$ are overlaid. Similar to the CaT equivalent width, there is a clear difference in color between Pic~II members and the foreground stars.
\label{fig:summary}}
\end{figure*}

\subsection{Updated Morphological Parameters}
\label{section:structure}

DELVE DR3 photometry extends roughly a magnitude fainter than the discovery analysis \citep{DrlicaWagner2016ApJ...833L...5D} and will provide significantly improved constraints on the structural and morphological parameters of Pic~II.
We use the \texttt{ugali} toolkit \footnote{\url{https://github.com/DarkEnergySurvey/ugali}} \citep{Bechtol2015ApJ...807...50B, DrlicaWagner2015ApJ...813..109D, DrlicaWagner2020ApJ...893...47D} to update the morphological parameters.
Briefly, \texttt{ugali} simultaneously fits the structural and stellar population parameters of a UFD. 
The structural component is fit with a \citet{Plummer1911MNRAS..71..460P} profile and the free parameters are: centroid coordinates ($\alpha_{\rm J2000}$, $\delta_{\rm J2000}$), semi-major axis ($a_h$), ellipticity ($\epsilon$), position angle (P.A.) of the major axis (defined east of north).
The magnitude and colors are fit with  \texttt{PARSEC} stellar isochrones  
\citep{Bressan2012MNRAS.427..127B, Chen2014MNRAS.444.2525C, Tang2014MNRAS.445.4287T, Chen2015MNRAS.452.1068C} with the following free parameters: distance modulus $(m-M)_0$, age ($\tau$), and metallicity ($Z_{\rm phot}$). The model lastly includes a parameter for the stellar richness ($\lambda$), which normalizes the total number of stars in the system \citep{DrlicaWagner2020ApJ...893...47D}. We use  \texttt{emcee} \citep{ForemanMackey2013PASP..125..306F} to simultaneously fit all parameters in the model. The posteriors from the \texttt{ugali} fits are summarized in Table~\ref{table:properties}.

We find an azimuthally averaged half-light radius of $R_{1/2}\sim32 ~{\rm pc}$, an ellipticity of $\epsilon\sim0.22$, an absolute $V$-band magnitude of $M_V\sim -2.6$, and a distance modulus of $(m-M)_0=18.45$ or $D_{\odot}=45~{\rm kpc}$.
To compute the $V$-band magnitude of Pic~II, we convert from $g,r$ magnitudes using the relation in \citet{DES2021ApJS..255...20A}.
Compared to the discovery paper \citep{DrlicaWagner2016ApJ...833L...5D}, the centroid is offset by  $\approx17^{\prime\prime}$, the size is 25\% smaller ($R_{1/2}\sim 32~{\rm pc}$ vs $R_{1/2}\sim43~{\rm pc}$),  the luminosity is 30\% lower ($M_V\sim -2.6$ versus $M_V\sim-3.2$), and shape and orientation are  better constrained. 
Our updated morphological parameter measurements are broadly consistent within the uncertainties of the discovery analysis \citep{DrlicaWagner2016ApJ...833L...5D}.
With our updated size and luminosity measurement, Pic~II remains  larger than known globular clusters and ambiguous systems and has properties  similar to other MW and LMC UFDs.

\subsection{Spectroscopic Membership Classification}
\label{section:member}

\begin{figure}
\centering
\includegraphics[width=0.95\columnwidth]{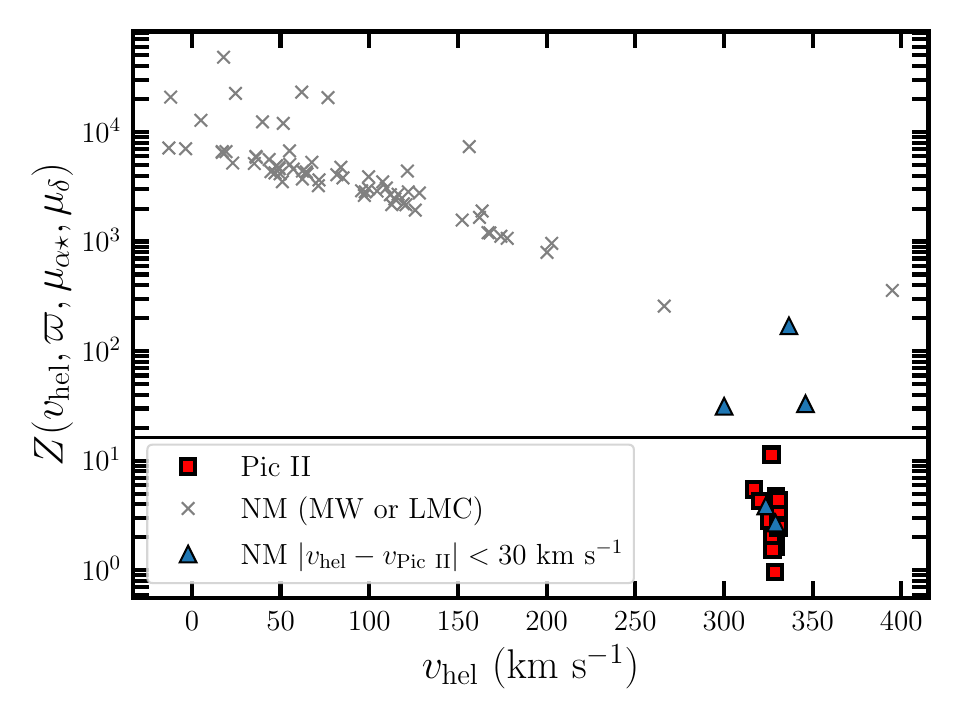}
\caption{
Membership score \citep{Tolstoy2023A&A...675A..49T} for stars in our spectroscopic sample. The horizontal line at 16.2 corresponds to a $3\sigma$ cut for selecting members. The colors and markers are the same as Figure~\ref{fig:summary}.
\label{fig:zscore}}
\end{figure}

We  assess the membership of each star based off its kinematics, chemistry, and photometric properties. We note that ``background'' stars here can refer to either foreground MW stars or  LMC stars at a similar distance.
The properties of the spectroscopic sample are summarized in Figure~\ref{fig:summary}.  
We highlight our final spectroscopic sample, along with clear background stars, and background stars with a velocity similar to Pic~II but classified as non-members.
The velocity distribution of our spectroscopic sample has two peaks, a wide peak at $\sim50~{\rm km~s^{-1}}$ which we associate with the MW and a second narrow peak at $v_{\rm hel} \sim 330~{\rm km~s^{-1}}$ which we associate  as Pic~II. 
The majority of the stars closest to Pic~II match the $v_{\rm hel} \sim 330~{\rm km~s^{-1}}$ peak and the  velocity association is further confirmed by selecting on proper motion.
The proper motion selection can be done with the membership from  {\it Gaia} proper motion based mixture models  \citep[membership probability $>0.1$;][]{Battaglia2022A&A...657A..54B, Pace2022ApJ...940..136P} or a  proper motion and parallax selection such as the membership score ($Z_3 < 13.2$) from \citet{Tolstoy2023A&A...675A..49T}.
All proper motion selections uncover the same velocity peak which confirms the  Pic~II systemic velocity.

While the majority of the background peaks at $50~{\rm km~s^{-1}}$ there is a tail beyond $\sim 200~{\rm km~s^{-1}}$ that we associate with the LMC. The  velocity at the center of the LMC is  $v_{\rm hel}= 262\pm3.4~{\rm km~s^{-1}}$ \citep{vanderMarel2002AJ....124.2639V} and at the position of Pic~II, the LMC velocity is $v_{\rm hel}\approx 300~{\rm km~s^{-1}}$ \citep[see Figure 3 of ][]{vanderMarel2014ApJ...781..121V}. 
We note the  excess of stars near the Pic~II systemic velocity that are non-members and these stars have kinematics that are consistent with the broad LMC distribution. 
As Pic~II is a likely LMC satellite (see Section~\ref{section:lmc}), it is not surprising to see some overlap with the LMC foreground.
We note that LMC stars have been found in the foreground of other dwarf galaxies including Carina \citep{Munoz2006ApJ...649..201M} and Carina~II \citep{Li2018ApJ...857..145L}.

Our final membership selection also considers metallicity, derived both from the spectroscopic CaT equivalent width and the CaHK photometry, and the location on an old, metal-poor isochrone \citep[age=12.5 Gyr and Z=0.0001]{Dotter2008ApJS..178...89D}.
The lower-middle panel of Figure~\ref{fig:summary} compares  the absolute V-band magnitude with the spectroscopic CaT equivalent width. Overlaid are  lines of constant [Fe/H] from \citet{Carrera2013MNRAS.434.1681C}.
The spectroscopic members have much lower equivalent widths than the background stars.
We present a color-color diagram in ${\rm CaHK} - g - 0.9~(g-i)$ vs. $g-i$ space in the lower-right hand panel of Figure~\ref{fig:summary}, which has been established to define axes where metal-poor and metal-rich stars separate from each other (e.g., \citealt{Keller2007PASA...24....1K, Starkenburg2017MNRAS.471.2587S, Chiti2020ApJ...891....8C, Huang2022ApJ...925..164H}).
As expected, we find that our spectroscopic members of Pic~II appear more metal-poor than MW halo stars in this plot, adding confidence to their membership.
For completeness, we note that the $g,i$ photometry in this metallicity-sensitive color-color plot is from DELVE DR2, and that iso-metallicity contours from a grid of Turbospectrum-generated synthetic spectra at log\,$g = 2$\,dex   are overplotted \citep{Chiti2020ApJ...891....8C, Barbosa2025arXiv250403593B}.

To quantitatively assess membership, we compute the membership score \citep{Tolstoy2023A&A...675A..49T}.
The membership score is effectively a 3-$\sigma$ selection around the systemic properties with either  proper motion and parallax, $Z_3$, or proper motion, parallax, and velocity, $Z_4$.
Members are identified within $Z_4<16.3$ (the 3$\sigma$ limit for a 4-dimensional $\chi^2$ distribution) and  we show the membership score of all stars  in Figure~\ref{fig:zscore}.  All members previously identified satisfy this quantitative selection.

Of the non-members within $30~{\rm km~s^{-1}}$ of the Pic~II velocity (green points in Figure~\ref{fig:summary}), five stars are clear astrometric non-members and two stars have a metallicity (either CaT EW or CaHK colors) that is much larger than Pic~II.
Both of the metallicity outlier stars are consistent with the velocity and proper motion of Pic~II and have $Z_4<16.3$. The first, (ID=10286300169398), has ${\rm [Fe/H]_{\rm CaT}}=-1.53\pm0.15$ and is 1.2 dex more metal-rich than any of the Pic~II members. This star does not have a ${\rm [Fe/H]_{\rm CaHK}}$ in the MAGIC catalog but is located in the CaHK color space near this metallicity. Lastly, we note this star is located on the old, metal-poor stellar isochrone. While a [Fe/H] $\sim -1.5$ star in a UFD is not uncommon or implausible a large metallicity gap seems unlikely and we consider this star a likely LMC interloper. The second, (ID=10286300169327), does not have a CaT EW measurement and has  ${\rm [Fe/H]_{\rm CaHK}}=-1.2$. This star is redder than the stellar isochrone. As this star is even more metal-rich it is even less likely to be a Pic~II member. 
While both these stars are more metal-poor than the bulk of LMC, they are consistent with the LMC halo metallicity distribution \citep[e.g.,][]{Borissova2006A&A...460..459B}.
In total, we identify 13 Pic~II members, which includes two horizontal branch stars.

The spatial distribution of the spectroscopic member sample in Figure~\ref{fig:summary} is quite asymmetric and nearly all the members are located on the southern side of Pic~II (all  spectroscopic members are at $y<1\arcmin$). 
Obtaining this distribution by chance is unexpected and there  are more successful velocity measurements on the northern side ($y>0$; N=63)  than the southern side ($y<0$; N=49) of Pic~II. 
This asymmetry is partly caused by the presence of a nearby relatively bright star ($G\sim 9.4$) that is located 2\arcmin~from the center of Pic~II and  there are only a few sources within $\approx1$\arcmin~ of this star in  our photometric catalogs. 
Two of our five masks utilize {\it Gaia} astrometry to target high priority targets missing from past observations.
Neither of these masks targeted the northern central side of Pic~II.
There are additional unobserved high-probability {\it Gaia} candidates \citep{Battaglia2022A&A...657A..54B} that are outside our {\it Gaia} target selection (three high probable candidates are near the bright star) and  if these stars were confirmed  as members, the spectroscopic sample would become more symmetric.
Overall, we suspect that our spectroscopic target selection and the presence of a nearby bright star is the cause of at least some the asymmetric spectroscopic member distribution.

\subsection{Constraints on Binarity}

Almost half of the  Pic~II spectroscopic members have repeat velocity measurements. 
Of the 6 members with repeat measurements, only one star, ID=10286300001343\footnote{Gaia DR3 source\_id=5480252276831909376}, shows clear evidence for velocity variations.  
From the repeat measurements, this  star has a reduced $\chi^2$ of $\approx63$ with a probability of $P(\chi^2, n=4)\approx10^{-13}$ that the measurements are drawn from a constant velocity distribution and it is highly likely to be a binary star.
All other stars with repeat measurements do not show evidence for velocity variation. 
Due to the low number of epochs (there are four velocity measurements but 2 epochs are  1 day apart), we are not able to constrain any  binary orbital parameters.

\subsection{Stellar Kinematics}
\label{section:stellar_kinematics}

We assume  the  velocity distribution follows a Normal distribution and use an unbinned likelihood  to compute the systemic velocity and velocity dispersion \citep[e.g.,][]{Walker2006AJ....131.2114W}.
We use a Jeffreys prior for the velocity dispersion ($-2\leq \log_{10}\sigma_v \leq1.7$), and from the 13 spectroscopic member sample we find  $\overline{v}_{\rm hel}=326.9\pm1.1~{\rm km~s^{-1}}$ and $\sigma_v =3.5_{-0.9}^{+1.1}~{\rm km~s^{-1}}$.
If we assume a linear prior in the velocity dispersion ($0\leq \sigma_v \leq 20$) we find $\overline{v}_{\rm hel}=327.0\pm1.2~{\rm km~s^{-1}}$  and $\sigma_v=3.8_{-0.9}^{+1.3}~{\rm km~s^{-1}}$.
Including or excluding the binary star does not affect  our inferred stellar kinematics  due to the large velocity error for the  combined velocity measurement for this star ($v_{\rm hel}=329.1 \pm	7.5~{\rm km~s^{-1}}$). 

To assess the significance of resolving the velocity dispersion, we compute the Bayes factor ($B$) comparing a free velocity dispersion model and zero velocity dispersion model. The Bayes factor is the ratio of Bayesian evidence between two models and is a commonly utilized model selection tool \citep[see][ for a review]{Trotta2008ConPh..49...71T}. We use the \texttt{pocoMC} package \citep{Karamanis2022JOSS....7.4634K, Karamanis2022MNRAS.516.1644K} to sample the posterior distributions and compute the Bayesian evidence for both models. 
We find $\ln{B}=12.1$ strongly favoring the non-zero dispersion model\footnote{The Bayes factor drops  to 11.9 if we compare to the stellar only velocity dispersion of $0.18~{\rm km~s^{-1}}$ which still strongly favors the velocity dispersion model. For the linear prior we find $\ln{B}=12.4$ comparing to the zero dispersion model.}. We use  Jeffreys' scale to interpret the Bayes factor where values of $\ln{B}<1$, 1-2.5, 2.5-5, $>5$ correspond to inconclusive, weak, moderate, and strong evidence in favor of one model, respectively (negative value favor the other model) \citep{Trotta2008ConPh..49...71T}. Overall, we find strong evidence for a non-zero velocity dispersion.

To assess the robustness of the velocity dispersion to outliers,  we perform a jackknife test. This test recomputes the velocity dispersion by excluding one star at a time.
There are two stars (ID = 10286300169004, 10286300071017) that when excluded decrease the velocity dispersion to $\sim2.5~{\rm km~s^{-1}}$ from $\sim3.5~{\rm km~s^{-1}}$ and if both are removed the velocity dispersion becomes marginally unresolved (the posterior peaks at $\sim1.2~{\rm km~s^{-1}}$ but there is a tail in the posterior distribution consistent with zero velocity dispersion).
Neither star has a repeat velocity measurement and both are likely members given either {\it Gaia} astrometry or a metallicity measurement\footnote{10286300071017 does not have a IMACS CaT measurement due to low signal-to-noise nor a CaHK [Fe/H] measurement due to the lack of i-band photometry.  However, when g-r color is used in place of g-i color for the CaHK color-color diagram (i.e., in the lower-right hand panel of Figure~\ref{fig:summary}), 10286300071017 clusters with the other Pic~II members and is consistent with the Pic~II metallicity.}. 
It is possible that either or both are binary stars but  further radial velocity measurements are required to test binarity. 

Unresolved binary stars can inflate the inferred velocity dispersion \citep[e.g.,][]{Minor2010ApJ...721.1142M, McConnachie2010ApJ...722L.209M, Buttry2022MNRAS.514.1706B}.  To assess the impact the identified binary could have had on the stellar kinematics without our multi-epoch data, we recompute the velocity dispersion using the first epoch velocity measurement instead of the combined velocity ($v_{\rm first~epoch} = 346.8\pm2.8~{\rm km~s^{-1}}$). The binary-influenced stellar kinematics are $\overline{v}_{\rm hel}=328.1\pm1.8~{\rm km~s^{-1}}$  and $\sigma_v=6.3_{-1.3}^{+1.8}~{\rm km~s^{-1}}$. While the inferred velocity dispersion is almost twice as large, the binary star may not have been considered a member as it  is a  4.5$\sigma$ velocity outlier (with $\sigma_v=3.5~{\rm km~s^{-1}}$). Without the multi-epoch velocity data the velocity dispersion of Pic~II could have been inflated similar to Bo\"{o}tes~II and Triangulum~II \citep{Kirby2017ApJ...838...83K, Bruce2023ApJ...950..167B}.

We compute the  dynamical mass and mass-to-light ratio of Pic~II with the mass estimator from~\citet{Wolf2010MNRAS.406.1220W}. We note this estimator computes the dynamical mass within the 3D half-light radius ($r_{1/2}$) which is slightly larger than the projected half-light radius for a Plummer profile ($r_{1/2}\approx 4/3 R_{1/2}$).
From the  velocity dispersion measurement here and the half-light radius, distance, luminosity measurements in Section~\ref{section:structure}, we infer  $M_{1/2} =3.6_{-1.6}^{+2.9} \times 10^5~{\rm M_{\odot}}$ and $M/L = 760_{-420}^{+910}\,{\rm M_{\odot}/L_{\odot}}$. The high dynamical mass-to-light ratio implies that Pic~II is a dark matter-dominated system.  

A velocity gradient is a clear sign of disequilibrium and has been searched for in many UFDs \citep[e.g.,][]{Martin2010ApJ...721.1333M, Collins2017MNRAS.467..573C, Li2018ApJ...866...22L, Ou2024ApJ...966...33O}. 
The velocity gradient model adds  two additional parameters, the magnitude of the velocity gradient ($v_g$), and its direction ($\theta_g$). 
With our stellar-kinematic sample we measure an upper limit of $v_g < 1.3~{\rm km~s^{-1}~arcmin^{-1}}$ for the velocity gradient (95\% credible level).
With the majority of the spectroscopic sample located  on the southern side of Pic~II, the small sample size, and the angular extent of the data, measuring a non-detection is not surprising.

\subsection{Metallicity Dispersion}
\label{section:metallicity_dispersion}

We apply the same likelihood analysis to determine the mean metallicity and metallicity dispersion of Pic~II  assuming that the metallicity distribution of Pic~II is a Normal distribution (with a Jeffreys prior for the metallicity dispersion). From the nine RGB  members with good quality CaT measurements, we obtain $\overline{{\rm [Fe/H]}} = -2.99 \pm0.06$ and $\sigma_{\rm [Fe/H]} <0.18$ (95\% credible level). 
Using a linear prior instead for the metallicity dispersion, we find $\sigma_{\rm [Fe/H]} < 0.27$ (95\% credible level). 

To verify our results, we apply the same analysis to the photometric CaHK metallicity.
There are  ten RGB spectroscopic members with photometric metallicity measurements and  we determine the metallicity distribution to be $\overline{{\rm [Fe/H]}}_{\rm CaHK} = -2.98 \pm0.10$ and $\sigma_{\rm [Fe/H]_{\rm CaHK}} <0.35$ (95\% credible level). 
The [Fe/H] measurements with both spectroscopic CaT and photometric CaHK agree and infer that Pic~II is an extremely metal-poor system with a small (unresolved) metallicity dispersion.

We apply the same Bayes factor significant test to the metallicity to assess whether the metallicity dispersion is resolved. For linear priors, we find $\ln{B}=-1.7,~-0.4$ for the CaT and CaHK [Fe/H] measurements, respectively, and for Jeffreys priors we find $\ln{B}=-0.5,~0.1$ for the CaT and CaHK [Fe/H] measurements, respectively. In general, we do not find evidence for a non-zero metallicity dispersion and either a larger sample and/or more precise metallicity measurements are required to resolve the dispersion. 

As discussed in Section~\ref{section:member}, there are two more meta-rich stars that have a velocity and proper motion consistent with Pic~II. If we include the more metal-poor star of the two (ID=10286300169398; ${\rm [Fe/H]_{CaT}}\sim -1.5$) in the metallicity distribution calculation, we find $\overline{{\rm [Fe/H]}}_{\rm CaT} = -2.76_{-0.17}^{+0.18}$ and $\sigma_{\rm [Fe/H]_{\rm CaT}}=0.49_{-0.11}^{+0.16}$. These values are consistent with other UFD galaxies. Due to the large metallicity gap between this star and the current most metal-rich Pic~II star ($\sim -2.7$), we do not consider this star a member. 

The brightest member star in Pic~II has a CaT metallicity of ${\rm [Fe/H]}= -3.08\pm0.08$ and is further observed with high-resolution spectroscopy discussed in Section~\ref{section:abundances}.

\subsection{Systemic  Proper Motion}
\label{section:proper_motion}

As shown in Figure~\ref{fig:summary}, all  12 spectroscopic members with {\it Gaia} astrometry cluster nicely around $(\mu_{\alpha\star}, \mu_{\delta} ) \sim (1,1)~{\rm mas~yr^{-1}}$. 
From the 11 members with good quality {\it Gaia} astrometry, we calculate the systemic proper motion of Pic~II to be $\overline{\mu}_{\alpha\star} = 1.18\pm0.05 ~{\rm mas~yr}^{-1}$ and $\overline{\mu}_{\delta} = 1.16\pm0.05 ~{\rm mas~yr}^{-1}$ with the likelihood in \citet{Pace2019ApJ...875...77P}.
Our measurement agrees with literature  {\it Gaia} DR3 proper motion measurements \citep{McConnachie2020RNAAS...4..229M, Pace2022ApJ...940..136P, Battaglia2022A&A...657A..54B}.
We note that one member\footnote{Gaia DR3 \texttt{source\_id}=5480248011928003968}  is just outside our selection for good quality astrometry (\texttt{astrometric\_excess\_noise\_sig}$=2.2$).  While this star is excluded for our default measurement, including or excluding this star does not affect our systemic proper motion results.

\section{Discussion}
\label{section:discussion}

\subsection{The Nature of Pic~II}
\label{section:nature}

\begin{figure}
\centering
\includegraphics[width=0.95\columnwidth]{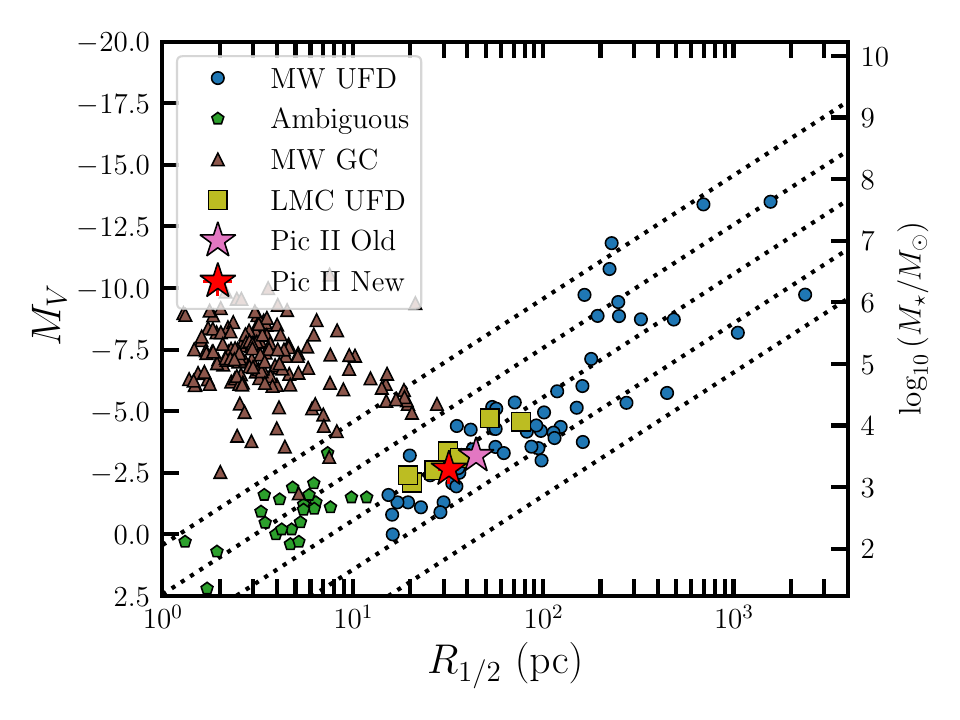}
\caption{Size ($R_{1/2}$) versus  absolute magnitude ($M_V$) comparison for MW satellites including dwarf galaxies, globular clusters, and ambiguous systems. We highlight LMC UFDs  and the new and old Pic~II measurements.  Contours of  constant surface brightness are indicated with dotted lines at $24,26,28,30,32~{\rm mag~arcsec^{-2}}$.
}
\label{fig:size_luminoisity}
\end{figure}

\begin{figure}
\centering
\includegraphics[width=0.95\columnwidth]{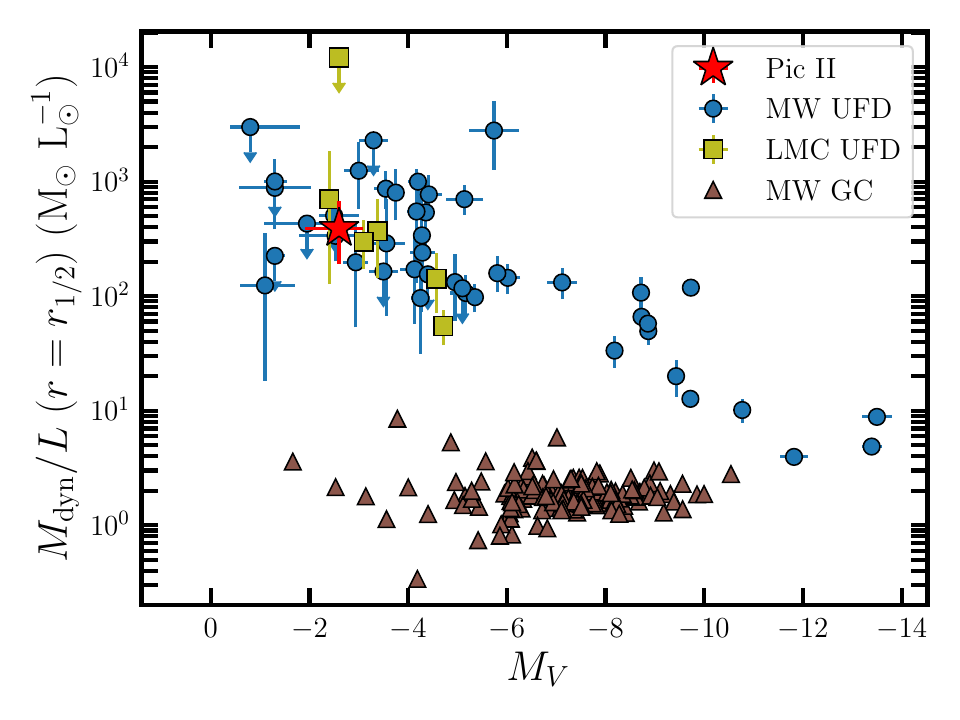}
\caption{Absolute V-band magnitude vs dynamical mass-to-light ratio comparing Pic~II (red star) to MW (blue circle) and LMC (gold squares) UFDs.
The large mass-to-light ratio of Pic~II is similar to other UFDs and Pic~II is a dark matter-dominated dwarf galaxy. 
}
\label{fig:mass_to_light}
\end{figure}

\begin{figure}
\centering
\includegraphics[width=0.95\columnwidth]{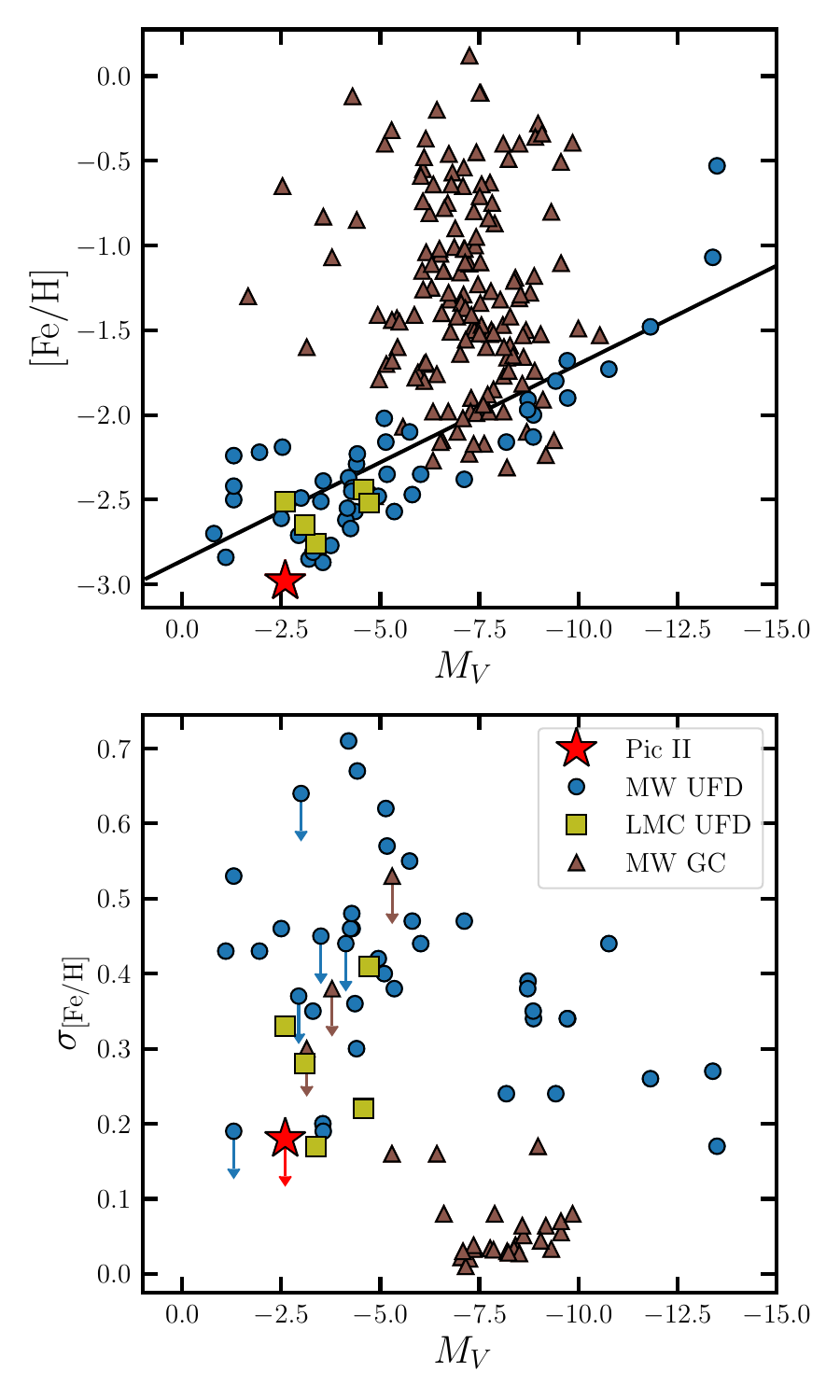}
\caption{Absolute V-band magnitude vs metallicity ([Fe/H], top panel) or metallicity dispersion ($\sigma_{\rm [Fe/H]}$, bottom panel) comparing Pic~II (red star) to MW (blue circle) and LMC (gold squares) UFDs and MW globular clusters (brown triangle). 
The black line is the stellar mass-stellar metallicity relation \citep{Kirby2013ApJ...779..102K, Simon2019ARA&A..57..375S}.
Pic~II is notably one of the most metal-poor UFD known.
}
\label{fig:metallicity_compare}
\end{figure}

Based on its size and luminosity, Pic~II was considered an UFD galaxy candidate in the discovery analysis \citep{DrlicaWagner2016ApJ...833L...5D}.
Following the definition of a galaxy from \citet[][i.e., the presence of a dark matter halo]{Willman2012AJ....144...76W}, our spectroscopic follow-up of Pic~II and measurement of the large dynamical mass-to-light ratio ($\sim700$) confirms the dwarf galaxy classification. 
While a dynamical classification is the most straightforward, as the resolved dispersion is dependent on the membership and velocity of 2 stars in  the 13 star member sample, it is worth exploring  the other  classification criteria.

A metallicity spread is indirect evidence for a presence of a dark matter halo massive enough to retain supernova ejecta to enable self-enrichment of a galaxy. 
The metallicity dispersion is not resolved with either CaT or CaHK [Fe/H] and this classification method is inconclusive for Pic~II. 
The mean metallicity of Pic~II is more metal-poor ($\overline{\rm [Fe/H]}\sim-3$) than any intact MW globular cluster  \citep{Harris1996AJ....112.1487H} and  is well below the globular cluster  metallicity floor  at [Fe/H]$\sim-2.5$ \citep[e.g.,][]{Beasley2019MNRAS.487.1986B}.  
While there are some  globular clusters stellar streams below the metallicity floor  \citep[e.g., Phoenix and C-19;][]{Wan2020Natur.583..768W, Martin2022Natur.601...45M}  as well as a M31 globular cluster  \citep{Larsen2020Sci...370..970L}, the MW lacks any intact systems below this value. 
The updated size of Pic~II ($r_{1/2}\sim 32~{\rm pc}$) with DELVE DR3 photometry remains larger than the typical globular cluster \cite[typically $r_{1/2}\lesssim 5~{\rm pc}$ but there are some GCs with $r_{1/2}\sim5-20~{\rm pc}$;][]{Baumgardt2018MNRAS.478.1520B} and it is located in the dwarf galaxy locus in the size-luminosity plane (see Figure~\ref{fig:size_luminoisity}).

Another defining characteristic of UFDs is their  low neutron-capture element abundances from their inefficient star formation\footnote{Although there are some exceptions: Reticulum~II, \citep{Ji2016Natur.531..610J, Roederer2016AJ....151...82R};  Tucana~III, \citep{Hansen2017ApJ...838...44H, Marshall2019ApJ...882..177M}), and Grus~II \citep{Hansen2020ApJ...897..183H}.} \citep[e.g.,][]{Koch2008ApJ...688L..13K, Frebel2015ARA&A..53..631F, Ji2019ApJ...870...83J, Ji2020ApJ...889...27J}.
The neutron capture abundances for the brightest star in  Pic~II are consistent with this picture (Section~\ref{section:abundances}). 
We only obtain upper limits for Ba, Eu, and Sr; the Sr and Ba upper limits exclude MW halo neutron capture values.
Triangulum~II and Grus~I are examples of galaxies with unresolved or barely resolved velocity/metallicity dispersions, but whose low neutron-capture element content suggests they are (or once were) UFDs \citep{Kirby2017ApJ...838...83K, Venn2017MNRAS.466.3741V, Ji2019ApJ...870...83J}.
In contrast, all known globular clusters have moderately super-solar neutron-capture element abundances  \citep[e.g.,][]{Pritzl2005AJ....130.2140P}.

Lastly, we compare the  properties of Pic~II with the MW and LMC UFD population. In general, Pic~II has properties similar to the bulk MW+LMC UFD population.
As we show in the next section, Pic~II is highly likely to be associated with the LMC and should be considered part of the LMC population for LMC and MW UFD comparisons (\S~\ref{section:lmc}). 
Figure~\ref{fig:size_luminoisity} compares the size and luminosity of the MW satellites and  includes both UFDs and  MW GCs. 
We note that  at fixed $M_V$, the LMC UFDs are smaller on average than MW UFDs but there may be selection effects as the LMC and many of its satellites are at lower Galactic latitude with higher stellar background. In a more detailed statistical analysis  \citet{Richstein2024ApJ...967...72R} did not find any differences between the LMC and MW in the size-luminosity plane. 
Figure~\ref{fig:mass_to_light} shows the absolute V-band magnitude versus the dynamical mass-to-light ratio.  Pic~II follows the trend  that lower luminosity UFDs have larger dynamical mass-to-light ratios.

Figure~\ref{fig:metallicity_compare} shows the absolute V-band magnitude versus the metallicity and metallicity dispersion. 
Pic~II is among the most metal-poor UFD known and is roughly 0.12 dex more metal-poor than Eridanus~IV \citep{Heiger2024ApJ...961..234H}, the previous most metal-poor UFD from spectroscopic measurements\footnote{We note that Canes Venatici~II and Hydra~II have photometric CaHK metallicities  that are more metal-poor or around the same metallicity as Pic~II \citep{Fu2023ApJ...958..167F} but these photometric metallicity measurements are more metal-poor than previous spectroscopic measurements \citep{Kirby2013ApJ...779..102K, Kirby2015ApJ...810...56K}.}.
The LMC UFDs are more metal-poor than expected from the  stellar mass-metallicity relationship and Pic~II is 0.44 dex more metal-poor than expected from this relation.
Our metallicity  measurements do not resolve a metallicity dispersion for  Pic~II and our upper limit is smaller than most  UFD   measurements. Only Hor~I, Eridanus~IV,  Leo~VI, and Tucana~III have  metallicity dispersions around or smaller than 0.20 dex, similar  to the Pic~II upper limit  \citep{Koposov2015ApJ...811...62K, Simon2017ApJ...838...11S, Heiger2024ApJ...961..234H, Tan2025ApJ...979..176T}. The small metallicity dispersion of Pic~II may be due to its lower star formation efficiency relative to other UFDs.

We note that most dwarf galaxies contain RR Lyrae stars \citep[e.g.,][]{MartinezVazquez2023MmSAI..94d..88M, Tau2024AJ....167...57T} and we present an unsuccessful RRL search in Appendix~\ref{section:rrl}. We list the literature references for Figures~\ref{fig:size_luminoisity}-\ref{fig:metallicity_compare} in Appendix~\ref{section:appendix_references}.

In summary, based on its size, low metallicity, mass-to-light ratio, and chemical abundances, we conclude Pic~II is an UFD and has properties similar to the UFD population.

\subsection{Orbit and Connection to the Magellanic Clouds}
\label{section:lmc}

\begin{figure*}
\centering
\includegraphics[width=0.49\textwidth]{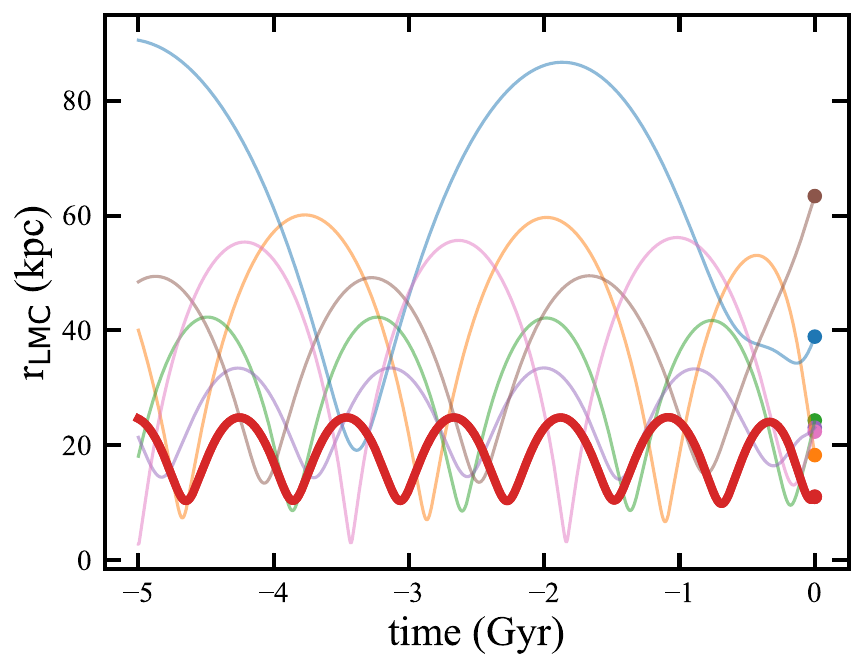}
\includegraphics[width=0.49\textwidth]{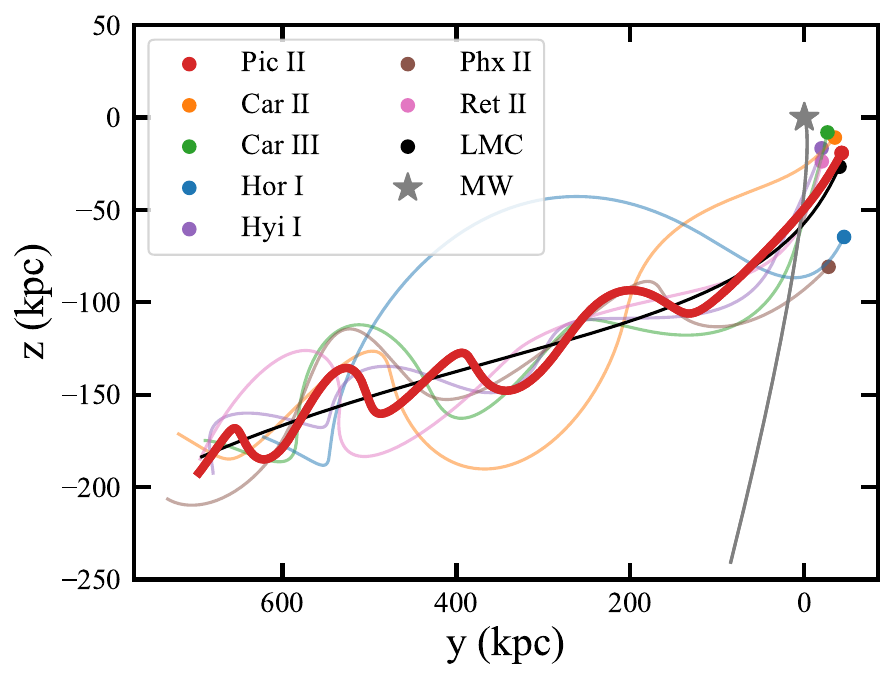}
\caption{
Example orbits of Pic~II and other  LMC associated UFDs relative to the LMC  and MW. {\bf Left}:  Lookback time versus relative distance to the LMC. 
Periodicity of plotted distances shows that the considered UFDs are gravitationally bound to the LMC in a given realization. {\bf Right}: Galactocentric coordinates (y vs z)   relative  to the current MW center. As the UFDs shown in this example are bound to the LMC, they move away from MW as a group over time.
\label{fig:pic2_orbit}}
\end{figure*} 

\begin{figure*}
\centering
\includegraphics[width=0.95\textwidth]{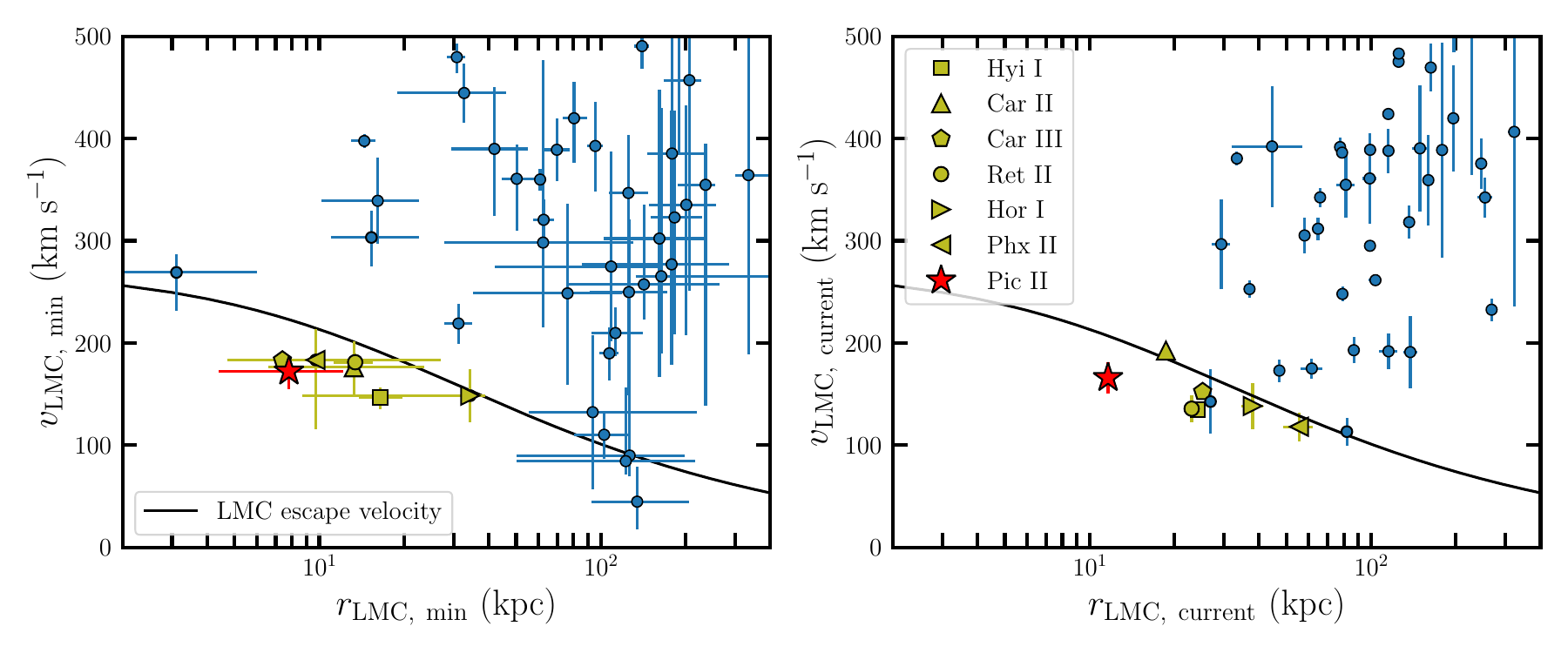}
\caption{
Distance and velocity relative to the LMC at each UFD's previous closest approach to the LMC (left) and current distance and velocity relative to the LMC (right). 
Both at closest approach and present moment, Pic II (red star) and other confirmed LMC UFDs (gold symbols) are under or very close to the LMC escape velocity curve, which is further evidence of their association with the LMC.
The MW UFD (blue circles) data is from \citet{Pace2022ApJ...940..136P}. 
The black line is the LMC escape velocity curve with $M_{\rm LMC} = 1.38 \times 10^{11}~M_{\odot}$. 
\label{fig:lmc_association}}
\end{figure*}

With a 3D distance of $\sim12$~kpc from the LMC, \citet{DrlicaWagner2016ApJ...833L...5D} suggested that Pic~II was likely accreted with the LMC and  gravitationally bound to the LMC. With the 3D motions measured in this work, we compute the relative 3D velocity between the LMC and Pic~II to be $166.0_{-15.8}^{+14.9}~{\rm km~s^{-1}}$ and an association is possible given the low relative velocity.
\citet{Jethwa2016MNRAS.461.2212J} built dynamical models to predict the phase space distribution of LMC and SMC UFD satellites. These models predict that Pic~II will have a Galactic standard of rest line-of-sight velocity of $15~{\rm km~s^{-1}} < v_{\rm gsr} < 175~{\rm km~s^{-1}}$ and Pic~II measured value $v_{\rm gsr}\sim105~{\rm km~s^{-1}}$ is in the middle of this distribution.

To verify a potential association, we compute the orbit of Pic~II in a combined MW and LMC potential.
We follow the methods of \citet{Erkal2020MNRAS.495.2554E} which treats the MW and LMC as individual particles that source their respective potentials and are able to move in response to each other.
This technique accounts for the MW reflex motion with respect to the infall of the LMC  \citep{Gomez2015ApJ...802..128G} and it includes the effect of dynamical friction of the MW on the LMC calculated with the prescription in \citet{Jethwa2016MNRAS.461.2212J}.
For the MW potential, we use the  \citet{McMillan2017MNRAS.465...76M}  potential, which includes an Navarro, Frenk, and White (NFW) halo \citep{Navarro1996ApJ...462..563N}, a stellar bulge, and four disks (thin, thick, HI, and H2).
To account for uncertainties in the MW potential, we sample over the posterior chains from the  \citet{McMillan2017MNRAS.465...76M} analysis.
For the LMC potential, we use a Hernquist profile with a total mass of $1.38 \pm 0.255 \times 10^{11}~{M_{\odot}}$ \citep[from][]{Erkal2019MNRAS.487.2685E} and a scale radius chosen to match the enclosed mass of $1.7\times10^{10}~M_{\odot}$ at 8.7 kpc from \citet{vanderMarel2014ApJ...781..121V}.
In addition to Monte Carlo sampling over the LMC mass measurement, we sample over the observed phase space uncertainties including the radial velocity \citep{vanderMarel2002AJ....124.2639V}, the proper motion \citep{Kallivayalil2013ApJ...764..161K}, and the  distance
\citep{Pietrzynski2019Natur.567..200P}.
The orbit of Pic~II is rewound for 5 Gyr or until the LMC passes its apocenter in the MW and LMC potential.
For more details see \citet{Erkal2020MNRAS.495.2554E}.

We use  our new Pic~II phase space measurements (Table~\ref{table:properties} and Section~\ref{section:properties}) sampled over the observational uncertainties to  compute the orbit of Pic~II in the presence of the MW and LMC.
In Figure~\ref{fig:pic2_orbit}, we show example orbits of Pic~II and LMC UFDs with respect to the LMC and MW in both lookback time and spatial location. In this realization, the considered UFDs are gravitationally bound to the LMC, so their distances relative to the LMC are periodic. They also move away from MW as a group over time in galactocentric coordinates.
Since we do not account for the evolution of the MW or the LMC potentials, we compute the pericenter and apocenter here at their global minimum and maximum with respect to the MW or LMC.
We find the pericenter and apocenter of Pic~II are $\sim39~{\rm kpc}$  and $\sim219~{\rm kpc}$  with respect to the MW and are $\sim8~{\rm kpc}$  and $\sim30~{\rm kpc}$ with respect to the LMC. 
\citet{DSouza2022MNRAS.512..739D} find that the most recent pericenter and apocenter can be reliably determined during backwards integration but previous pericenters or apocenters are more unreliable.

We estimate the probability of Pic~II being an LMC satellite by computing its energy relative to the LMC 5 Gyr ago (at the end of integration) following \citet{Erkal2020MNRAS.495.2554E} to determine whether they were energetically bound. 
If the LMC reaches the apocenter before 5 Gyr, we stop the rewinding and compute the energy at that time.
With this definition, we compute that Pic~II has a 91\% association probability with the LMC\footnote{\citet{Patel2020ApJ...893..121P} present an alternative LMC probability that is determined on whether the satellite was within the escape velocity of the LMC at its most recent approach to the LMC and we find similar results with this definition.}.
We show this visually in Figure~\ref{fig:lmc_association}, where we show the minimum distance between Pic~II and LMC compared to the relative velocity between Pic~II and the LMC along with the current values. 
We note that \citet{CorreaMagnus2022MNRAS.511.2610C} computed orbits of Pic~II in a LMC and MW potential but varied the line-of-sight velocity between -200 to 800 ${\rm km~s^{-1}}$ and found that at $v_{\rm hel}\sim330~{\rm km~s^{-1}}$ the probability of LMC association is near 100\% (in particular the range of 100 to 500 ${\rm km~s^{-1}}$ has high probability for LMC association).

Pic~II is the seventh LMC UFD to be confirmed with full phase space measurements and orbit modeling.
The spatial distribution of the LMC UFDs is asymmetric  and there are more LMC UFDs in the leading  orbit of the LMC (Car~II, Car~III, Hyi~I,  Pic~II, and Ret~II) than trailing the orbit (Hor~I and Phe~II).  
This spatial anisotropy is similar to  M31 \citep[e.g.,][]{McConnachie2006MNRAS.365..902M, Savino2022ApJ...938..101S}, where there are more satellites on the MW side of the satellite system compared to the far side. With the LMC system there are fewer confirmed UFDs overall and the system is being tidally heated by the MW.  The  satellite sensitivity is lower for more distant systems and the coverage for satellite searches is lower on the southern side of the LMC and this asymmetry may be due to selection effects.
This asymmetry does not appear long lived as the number of LMC satellites on the near/far sides changes multiple times during our orbital integration.
Of the known UFD candidates without spectroscopy, DELVE 2 \citep{Cerny2021ApJ...910...18C} is a candidate member of the LMC system based on its spatial location and proper motion \citep{Cerny2021ApJ...910...18C, CorreaMagnus2022MNRAS.511.2610C, Vasiliev2024MNRAS.527..437V} and it is located on the far side of the LMC.

\begin{figure*}
\centering
\includegraphics[width=0.49\textwidth]{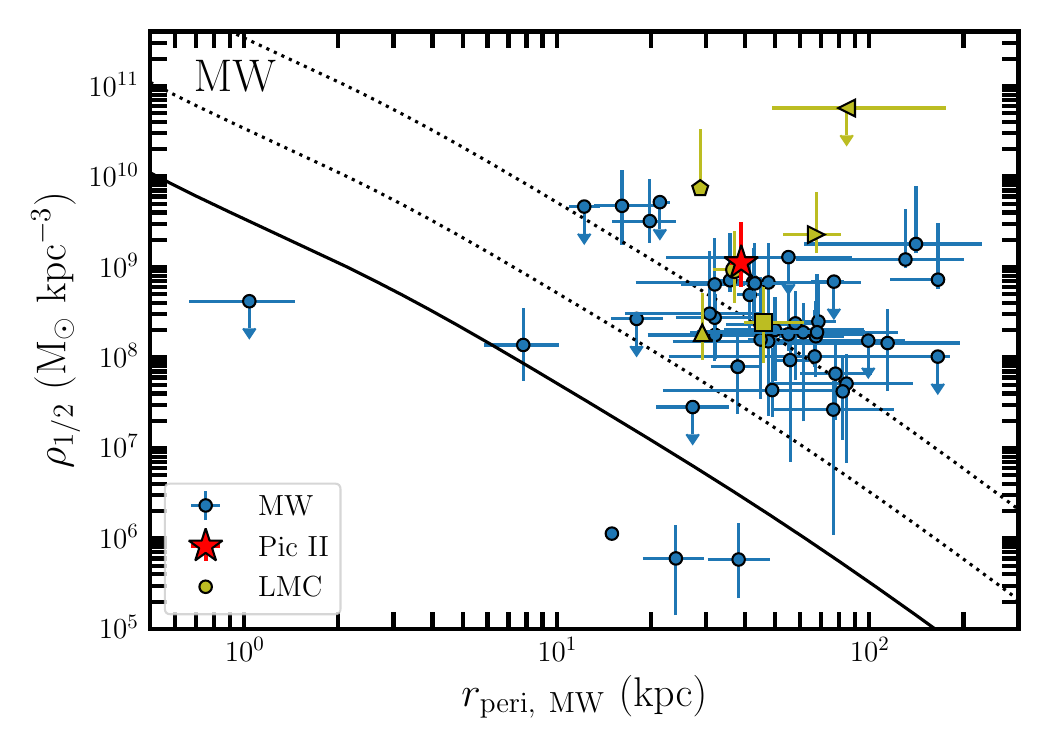}
\includegraphics[width=0.49\textwidth]{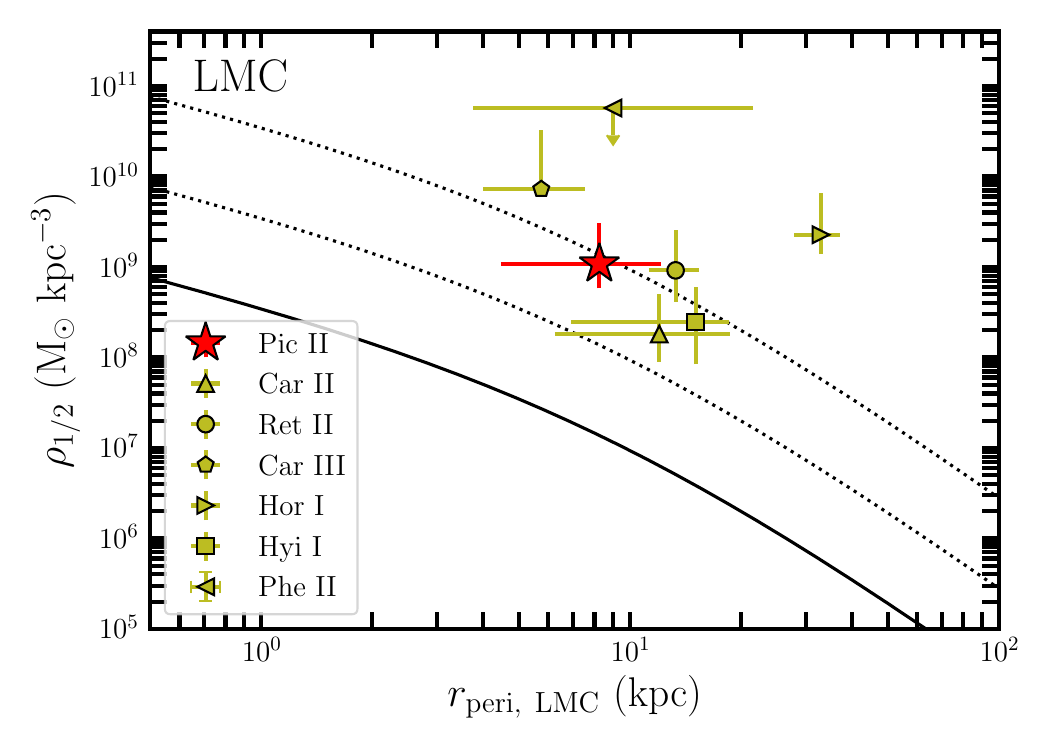}
\caption{
{\bf Left}: Orbital pericenter with respect to the MW versus average density with the half-light radius for  MW UFDs (blue points), LMC UFDs (gold points), and Pic~II (red point). The MW and LMC measurements are primarily from \citet{Pace2022ApJ...940..136P}  but include newer literature measurements. The solid black line is the twice the average MW density and this line denotes where the tidal radius at pericenter will be smaller or larger than the half-light radius. The dotted lines are 10 and 100 times this value.  {\bf Right}: Orbital pericenter with respect to the LMC versus average density with the half-light radius for Pic~II and other  LMC UFDs. The solid line is twice the average LMC density and the dotted lines are 10 and 100 times this value.
\label{fig:pericenter_density}}
\end{figure*}

In Figure~\ref{fig:pericenter_density}, we show the orbital pericenter versus average density within the half-light radius for Pic~II, MW UFDs, and LMC UFDs.  
Strong tidal disruption is expected when the average UFD density is less than twice the average MW or LMC density at its pericenter.
Pic~II is unlikely to be strongly stripped by the MW based on the Jacobii radius approximation. 
The right-hand panel of Figure~\ref{fig:pericenter_density} repeats the same exercise but with respect to the LMC instead of the MW.
The overall tidal influence the LMC exerts on its UFDs is smaller than the MW exerts on its UFDs due to its roughly order of magnitude smaller halo mass. 
The orbital pericenters with respect to the LMC of the LMC UFDs are generally smaller than the MW UFDs with respect to the MW and vary between $\approx 6-30~{\rm kpc}$ here.
While Pic~II has a  smaller pericenter ($\sim 8 ~{\rm kpc}$) with respect to the LMC compared to the MW, its tidal radius is still much greater than its half-light radius and strong tidal stripping is not expected.

To more quantitatively estimate the Pic~II tidal radius\footnote{See \citet{vandenBosch2018MNRAS.474.3043V} for an overview of the different tidal radius definitions.} with respect to the LMC, we first compute  the tidal radius  with $r_t = r_{\rm peri} \left(m_{UFD}/2 M_{\rm host} \right)^{1/3}$ which assumes the host has a flat rotation curve and the satellite is on a circular orbit \citep{King1962AJ.....67..471K}.
For the tidal radius computations, we assume that the mass distribution of Pic~II follows an NFW profile with several different scale radius values (specifically $r_s=40,100,200~{\rm pc}$) and set the scale density based on matching the $M_{1/2}$ dynamical mass. We assume that the LMC mass follows a Hernquist profile as above. 
We find $r_t=330, 450, 590~{\rm pc}$ for $r_s=40,100,200~{\rm pc}$.  The smallest $r_s$ value is set by the size of Pic~II ($\sim32~{\rm pc}$).  These tidal radii are more than 10 times larger than the half-light radius. 

To account for the non-circular motion and extended mass profiles, we also compute the tidal radius using $r_t = \left[ (G~m_{\rm UFD})/(\Omega^2 - \frac{{\rm d}^2\phi_{\rm host}}{{\rm d}R^2})\right]^{1/3}$, where $\Omega$ is the  angular speed and $\phi$ is the LMC gravitational potential \citep{King1962AJ.....67..471K}.
Here we find smaller tidal radii with the same scale radii values, $r_t=190, 270, 340~{\rm pc}$ and the smallest value is approximately 5 times larger than the half-light radius.
We note that in both the FIRE \citep{Shipp2023ApJ...949...44S} and Auriga \citep{Riley2024arXiv241009144R, Shipp2024arXiv241009143S} simulations nearly all  dwarf galaxy satellites in MW-like galaxies are undergoing tidal stripping and contain tidal tails but these are commonly below current detection limits \citep{Shipp2023ApJ...949...44S}. However, we note that some disruption in simulations may be numerical \citep{vandenBosch2018MNRAS.474.3043V}.
We note that the actual tidal radius should also account for the MW and SMC but both will have a  smaller impact relative to the   LMC.

\subsection{Detailed abundances of an extremely metal-poor star}
\label{section:abundances}

\begin{figure*}
\centering
\includegraphics[height=16cm]{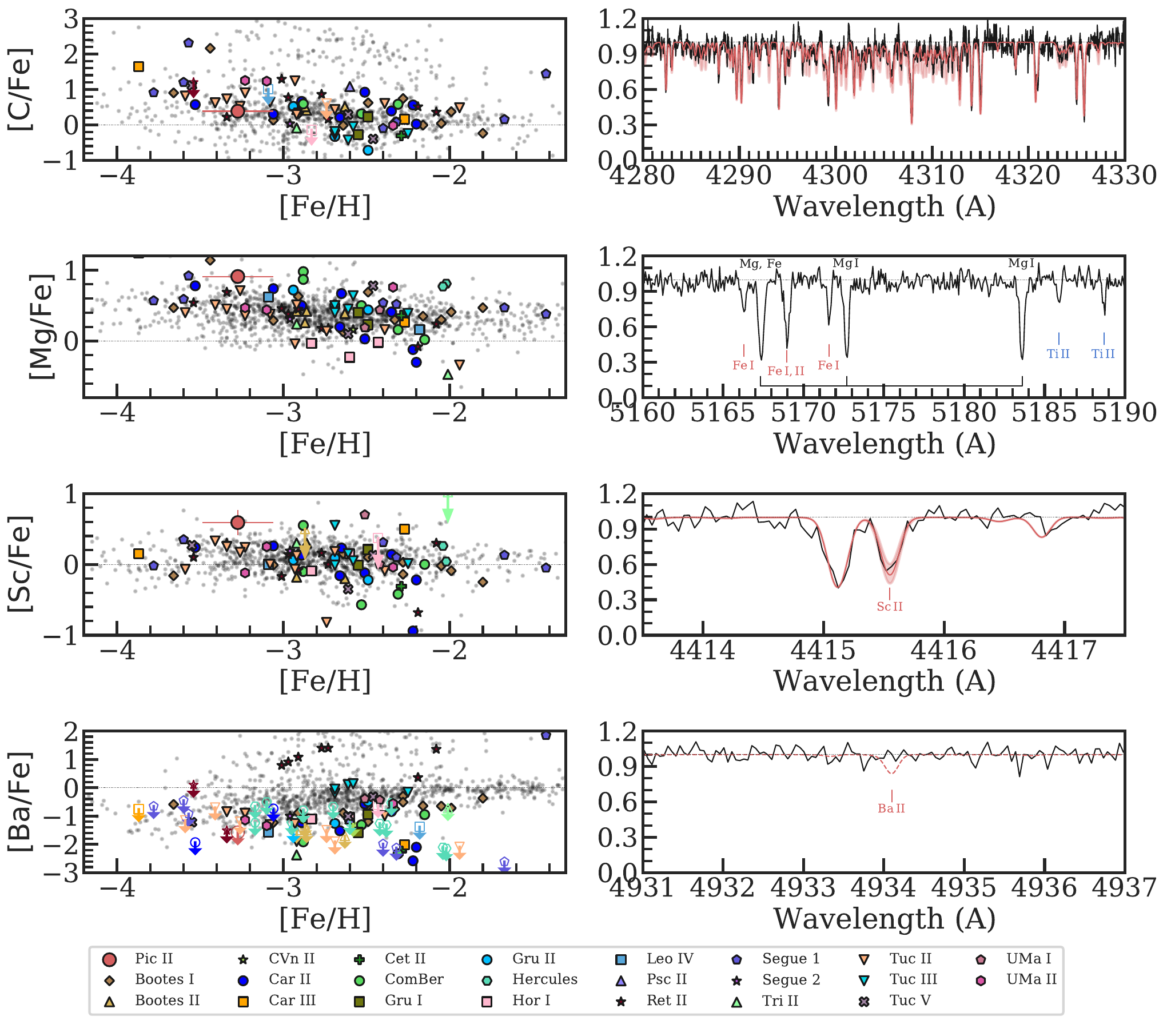}
\caption{
{\bf Left}: Detailed abundances of PicII-1, the brightest Pic~II star (large red circle) compared to halo stars (grey points, from \citealt{Abohalima2018ApJS..238...36A}) and other UFD stars (colored points).
{\bf Right}: Normalized MIKE spectra of the target star around the CH bands, Mg b triplet, a Sc line, and a Ba line. For CH and Sc, the synthesized fit shown as a red line, with $2\sigma$ abundance errors shaded. For Ba, a dashed red line shows the $5\sigma$ upper limit.
The literature references are listed in Appendix~\ref{section:appendix_references}}.
\label{fig:mike}
\end{figure*}

With $\mbox{[Fe/H]} = -3.3$, PicII-1 is one of the most Fe-poor stars known in any UFD. 
As such, the abundances of this star are closely connected to the nucleosynthetic yields of the first metal-free (Pop\,III) stars \citep[e.g.,][]{Ji2015MNRAS.454..659J}.
Still, most of its abundance ratios are quite similar to other metal-poor stars found in UFDs: the [$\alpha$/Fe] ratios are enhanced; most elements up to the iron peak are similar to halo stars at [Fe/H] $\sim -3$; and the neutron-capture element abundances are low compared to typical halo stars.
The abundances of PicII-1 are similar to one of the most metal-poor stars in the Coma Berenices UFD \citep[ComBer-S2 from][]{Frebel2010ApJ...708..560F}. ComBer-S2 is enhanced in Mg and Sc, deficient in Mn,  is not Carbon enhanced, and lacks neutron-capture elements \citep{Frebel2010ApJ...708..560F}.
In Figure~\ref{fig:mike}, we compare the [C/Fe], [Mg/Fe], [Sc/Fe], and [Ba/Fe] abundances to a representative  MW halo sample and abundance measurements from other UFDs.
The low neutron-capture element abundances lend additional confidence that Pic~II is a dwarf galaxy and not a star cluster (see Section~\ref{section:nature}) \citep{Ji2019ApJ...870...83J}.

Carbon-enhanced metal-poor stars are thought to be especially related to Pop\,III stars \citep[e.g.,][]{Frebel2015ARA&A..53..631F}.
About 40\% of halo stars with $\mbox{[Fe/H]} < -3$ have $\mbox{[C/Fe]} > 0.7$ \citep{Placco2014ApJ...797...21P, Arentsen2022MNRAS.515.4082A}, but it is still debated whether a similar trend is found in dwarf galaxies like Sculptor \citep[e.g.,][]{Salvadori2015MNRAS.454.1320S, Chiti2018ApJ...856..142C} and Sagittarius \citep[e.g.,][]{Limberg2023ApJ...946...66L, Sestito2024A&A...690A.333S}.
We measure this Pic~II star to have $\mbox{[C/Fe]}=-0.36$, and corrections for internal mixing in red giant evolution \citep{Placco2014ApJ...797...21P} suggest a natal abundance $\mbox{[C/Fe]} \approx +0.37$. This star is thus clearly \emph{not} significantly carbon-enhanced.
Including this star, there are now 22 stars across 9 UFDs that have $\mbox{[Fe/H]} < -3$ and carbon measurements: 3 in Segue\,1, \citep{Frebel2014ApJ...786...74F}; 2 in UMa II, \citep{Frebel2010ApJ...708..560F}; 3 in Bootes\,I, \citep{Norris2010ApJ...711..350N, Gilmore2013ApJ...763...61G, Frebel2016ApJ...826..110F}; 3 in Ret~II; \citep{Ji2016ApJ...830...93J, Roederer2016AJ....151...82R}; 6 in Tuc~II, \citep{Ji2016ApJ...832L...3J, Chiti2018ApJ...857...74C, Chiti2023AJ....165...55C}; 2 in Car II \citep{Ji2020ApJ...889...27J}; 1 in Car III \citep{Ji2020ApJ...889...27J}; and 1 in Tuc V \citep{Hansen2024ApJ...968...21H}.
13 (8) of these 22 stars are carbon-enhanced with $\mbox{[C/Fe]} > 0.7$ ($> 1.0$), which is  higher than but consistent with the stellar halo fraction, as shown in \citet{Ji2020ApJ...889...27J}.
We note that every one of these UFDs other than Pic~II has at least one carbon-enhanced star at $\mbox{[Fe/H]} < -3$, but with one star in Pic~II it would be premature to conclude that Pic~II does not contain any carbon-enhanced stars \citep[e.g.,][]{Koch2008ApJ...688L..13K, Gilmore2013ApJ...763...61G, Ishigaki2014A&A...562A.146I, Ji2016ApJ...832L...3J, Ji2019ApJ...870...83J, Ji2020ApJ...889...27J, Chiti2023AJ....165...55C, Waller2023MNRAS.519.1349W, Webber2023ApJ...959..141W, Hansen2024ApJ...968...21H}.

The $\alpha$-elements (magnesium, silicon, calcium, titanium) in this Pic~II star have super-solar ratios, similar to stars in all UFDs other than Hor~I.
\citet{Nagasawa2018ApJ...852...99N} found three stars in that system with $[\alpha/{\rm Fe}] \approx 0$. 
As Hor~I is very likely associated with the LMC \citep{Koposov2015ApJ...811...62K}, those authors hypothesized that the low $\alpha$-element abundances could be an indication of association with the LMC, possibly due to an overall lower star formation efficiency for galaxies forming near the LMC instead of the MW.
However, both Car~II \citep{Ji2020ApJ...889...27J} and Ret~II \citep{Ji2016ApJ...830...93J, Roederer2016AJ....151...82R}, two other LMC UFDs, do not show this trend.
Pic~II provides additional evidence against that hypothesis, as it is also clearly associated with the LMC but contains a star with enhanced [$\alpha$/Fe] ratios.
However, this Pic~II star has $\mbox{[Fe/H]} = -3.3$, significantly lower than the $\mbox{[Fe/H]} \sim -2.6$ stars studied so far in Hor~I.
This low-[Fe/H] star does not probe the late-time chemical evolution of the system, and so cannot really test the hypothesis that star formation efficiencies are different for LMC satellites.
Unfortunately, we currently have no $\alpha$-element constraints in Pic~II stars of higher [Fe/H].

Ratios of different $\alpha$-elements provide some constraint on the progenitor mass of a core-collapse supernova \citep[e.g.,][]{McWilliam2013ApJ...778..149M, Carlin2018ApJ...859L..10C, Ji2020ApJ...889...27J}. In particular, the yield of hydrostatically synthesized elements like Mg increases significantly with progenitor mass, while explosively synthesized elements like Ca are less affected by progenitor mass.
The detailed relation between [Mg/Ca] and progenitor mass is not straightforward, as Ca yields are affected by the SN explosion, but higher [Mg/Ca] ratios correspond to higher progenitor masses on average \citep[e.g.,][]{Heger2010ApJ...724..341H}.
We measure $\mbox{[Mg/Ca]} = 0.51 \pm 0.14$ in this star, a high ratio that suggests this Fe-poor star is predominantly (or perhaps only) enriched by relatively massive core-collapse supernovae.
Non-LTE corrections corrections\footnote{Online correction tools at \url{https://spectrum.inasan.ru/nLTE/} for Ca, \url{https://www.inspect-stars.com} for Mg, and \url{https://nlte.mpia.de} for both.} for this star result in small increases in Mg ($0.04-0.07$ dex, \citealt{Osorio2015A&A...579A..53O,Bergemann2017}) and moderate increases in Ca ($0.25-0.28$ dex, \citealt{Mashonkina2007,Mashonkina2016}), so the [Mg/Ca] ratio is 0.2 dex lower but still significantly enhanced.

\citet{Ji2020ApJ...889...27J} explored the evolution of [Mg/Ca] versus [Fe/H] for the UFD population and while most UFDs do not show a trend with [Fe/H], two LMC satellites (Car~II and Ret~II), show a declining trend with increasing [Fe/H]. 
The lowest metallicity star in Car~III also has high [Mg/Ca] but there are only two stars in Car~III with chemical abundance measurements.
This trend and its difference from other non-LMC UFDs could be evidence for environmental effects in the chemical evolution of a UFD \citep[see Section 4.3.3 of][]{Ji2020ApJ...889...27J}.
However, \citet{Hansen2024ApJ...968...21H}  recently showed that Tucana V, which does not appear to be associated with the LMC, also exhibits a similar [Mg/Ca] trend as the Magellanic satellites identified by \citet{Ji2020ApJ...889...27J}.
The star formation histories of LMC UFDs suggest they formed later \citep{Sacchi2021ApJ...920L..19S, Simon2023ApJ...944...43S, Durbin2025arXiv250518252D}.
While we cannot determine how [Mg/Ca] evolves with [Fe/H] in Pic~II, the high value of $\mbox{[Mg/Ca]}=0.51$ at $\mbox{[Fe/H]}=-3.3$ suggests Pic~II may follow the trend observed in other LMC UFDs.

The main other element of note is scandium. We measured $\mbox{[Sc/Fe]} = 0.59 \pm 0.18$ from 7 different lines, an unusually high value compared to other UFDs and halo stars at similar [Fe/H].
We fit synthetic spectra including hyperfine splitting for all Sc lines, the line-to-line abundance scatter is low, and variations due to stellar parameters are included in the uncertainty.
We also verify that our analysis method reproduces the Sc abundance of several other UFD and standard halo stars to within 0.1 dex.
The somewhat high Sc abundance in this star is thus real and significant, at least in 1D LTE (3D and NLTE effects for Sc\,II have not been investigated in metal-poor red giant stars, \citealt{Bergemann2014dapb.book..169B}).
The interpretation of high Sc is somewhat unclear, as standard models of core-collapse supernovae tend to underproduce Sc \citep[see discussion in][]{Nomoto2013ARA&A..51..457N, Curtis2019ApJ...870....2C}.
Two ways of increasing Sc production include the $\nu p$-process \citep{Frohlich2006PhRvL..96n2502F} or high-energy hypernova \citep{Nomoto2006NuPhA.777..424N}, but the other elements available in our Pic~II star do not constrain these processes.
We note that Zn tends to be produced with Sc in these models, but our spectrum provides only a weak 3$\sigma$ (5$\sigma$) limit of $\mbox{[Zn/Fe]} < +1.03\ (+1.30)$.
Overall, it seems that there could be substantial [Sc/Fe] variation across different UFDs \citep[e.g.,][]{Ji2020ApJ...889...27J, Chiti2023AJ....165...55C}, which may point to interesting variations in the first supernova explosions.

\subsection{Astrophysical J-factor and D-factor}
\label{section:j_factor}

The MW UFDs are among the most promising targets for searches of the products of dark matter annihilation or decay as they are nearby, highly dark matter-dominated, and nearly background free \citep[e.g.,][]{FermiLATCollaboration2015PhRvL.115w1301A, GeringerSameth2015PhRvD..91h3535G, McDaniel2024PhRvD.109f3024M, Circiello2025ApJ...978L..43C}.  
While some individual UFDs are excellent targets on their own,  new UFDs are also  useful in stacked joint analyses to boost any signal or exclusion limits \citep[e.g.,][]{FermiLATCollaboration2011PhRvL.107x1302A, GeringerSameth2011PhRvL.107x1303G} and we analyze Pic~II here for future dark matter indirect detection studies.

The astrophysical component of the annihilation or decay flux is known as the J-factor or  D-factor,  respectively.  
The J-factor is the integral of the dark matter density squared over the line-of-sight: $J(\theta)=\int \rho_{\rm DM}^2 {\rm d}\Omega{\rm d}l$. The D-factor is the linear analog: $D(\theta)=\int \rho_{\rm DM} {\rm d}\Omega{\rm d}l$.  Here, $\rho_{\rm DM}$ is the dark matter density profile and the integral is performed over the solid angle, $\Delta\Omega$, and along the line-of-sight, $l$. 
The standard procedure in MW dwarf spheroidal galaxies to  determine $\rho_{\rm DM}$ is to fit line-of-sight stellar kinematic data with dynamical mass models computed from the  spherical Jeans equations\footnote{Although other dynamical mass models have been used \citep[e.g.,][]{Hayashi2016MNRAS.461.2914H, Evans2016PhRvD..93j3512E}.} \citep[e.g.,][]{Strigari2008Natur.454.1096S, Bonnivard2015MNRAS.453..849B, Pace2019MNRAS.482.3480P}.

We follow \citet{Pace2019MNRAS.482.3480P} for our J-factor analysis and briefly summarize the procedure. 
For the spherical Jeans modeling, we assume the stellar density profile follows a Plummer distribution \citep{Plummer1911MNRAS..71..460P}, the dark matter profile is an NFW halo \citep{Navarro1996ApJ...462..563N}, and the stellar anisotropy is constant with radius.  We account for observational errors in the structural parameters and compare the line-of-sight velocity dispersion from Jeans equations solutions to observed stellar kinematics with  an unbinned likelihood \citep{Martinez2011ApJ...738...55M, GeringerSameth2015ApJ...801...74G}. 
For the J- and D-factor calculation, the extent of the DM halo is required. 
We use the tidal radius computed at the orbital pericenter with respect to the LMC as it is smaller than the MW tidal radius.
We fix $r_t=0.4~{\rm kpc}$ (or $r_t=$0\fdg51) following the discussion in Section~\ref{section:lmc}.
We note that few  J-factor measurements in the literature have computed the  tidal radius at the orbital pericenter as many measurements predate {\it Gaia} proper motion measurements.

We measure the J-factor to be: $\log_{10}{\left(J/ {\rm GeV^2 ~cm^{-5}} \right)}=18.15_{-0.53}^{+0.54},~18.33\pm0.53, 18.48\pm0.55$ for solid angles of $\theta=0\fdg1, 0\fdg2, 0\fdg5$ (the J-factor does not increase beyond the tidal radius of 0\fdg51).  
Similarly, the D-factor we measure is: $\log_{10}{\left(D/ {\rm GeV~cm^{-2}} \right)}=17.02\pm0.28, 17.45_{-0.31}^{+0.30}, 17.86_{-0.39}^{+0.33}$ for solid angles of $\theta=0\fdg1$, $0\fdg2$, $0\fdg5$ ($r_t=0.4~{\rm kpc}$).
The J-factor we measure  matches the dynamical J-factor scaling relations introduced in \citet{Pace2019MNRAS.482.3480P}, however, it does not match the predictions of the luminosity based scaling relations (the prediction is $\log_{10}{\left(J(0.5\degr)\right)}=18.9$). 
If the tidal radius is decreased to $r_t=0.3~{\rm kpc}$ (or $r_t=$0\fdg38), the J-factor and D-factor at smaller angles only slightly decrease whereas  the maximum angle  decreases to $\log_{10}{\left(J/ {\rm GeV^2 cm^{-5}} \right)}= 18.42_{-0.54}^{+0.53}$ and $\log_{10}{\left(D/ {\rm GeV~cm^{-2}} \right)} = 17.66_{-0.34}^{+0.31}$ for a solid angle of $\theta=0\fdg38$.  
The impact and decrease in the J-factor will be larger for smaller values of the tidal radius. 
We note that the D-factor is impacted more than the J-factor with smaller tidal radii.
Pic~II is a  good target for dark matter annihilation searches as  the J-factor is fairly large \citep[there are 15 MW dwarf galaxies larger in Figure~11 in][]{Heiger2024ApJ...961..234H} and it will be a useful addition in stacked analysis.

\section{Summary}
\label{section:summary}

We have presented the first stellar spectroscopy of the UFD candidate Pic~II. Our results are summarized as follows:

\begin{itemize}
    \item Using DELVE DR3 and the \texttt{ugali} package, we have updated the morphological parameters with data approximately 1 magnitude deeper than the discovery analysis. Our updated results find Pic~II to be slightly smaller ($R_{1/2}\sim 32~{\rm pc}$) and fainter ($M_V \sim -2.6$) than previous measurements. 
    \item We obtained 4 epochs of Magellan/IMACS medium resolution stellar spectroscopy and identified 13 radial velocity members. We determined the systemic heliocentric velocity of Pic~II to be $v_{\rm hel} = 326.9\pm1.1~{\rm km~s^{-1}}$ and resolve the velocity dispersion to be $\sigma_v=3.5_{-0.9}^{+1.1}~{\rm km~s^{-1}}$. 
    \item We measured both spectroscopic and photometric stellar metallicities ([Fe/H]) with Calcium Triplet equivalent widths and CaHK photometry and find good agreement. From spectroscopic [Fe/H], we found  $\overline{\rm [Fe/H]}=-2.99\pm0.06$ for Pic~II but are not able to resolve the metallicity dispersion ($\sigma_{\rm [Fe/H]}<0.18$; 95\% credible level).
    \item We determined detailed chemical abundances of 13 elements with a high resolution spectrum of the brightest Pic~II star  with Magellan/MIKE. Notably, the star is extremely metal-poor ([Fe/H]$=-3.3$), enhanced in $\alpha$ elements, and deficient in neutron-capture elements, similar to the abundances in other UFDs. 
    However, this star has an unusually high [Sc/Fe] ratio.
    \item The  dynamical mass-to-light ratio ($\sim760$), size, and low neutron-capture elements confirm that Pic~II is a dark matter-dominated UFD galaxy. 
    \item We computed the orbit of Pic~II in a combined MW and LMC potential and found that Pic II is bound to the LMC in 91\% of modeled orbits  and likely entered the MW system with the LMC. 
    The orbital parameters with respect to the LMC are $r_{\rm peri, \, LMC}=7.8^{+4.4}_{-3.4}~{\rm kpc}$ and $r_{\rm apo, \, LMC}=29.9^{+9.7}_{-6.0}~{\rm kpc}$. 
    Pic~II is the 7th UFD confirmed to be associated with the LMC  based on its measured phase space and orbital analysis. Pic~II is likely still bound to the LMC today.
    \item We computed the dark matter  properties of Pic~II in the context of dark matter indirect detection studies and measured the predicted annihilation and decay flux.  Pic~II has a fairly typical  J-factor of $\log_{10}{J(0\fdg5)}= 18.50\pm0.55$ and will be a useful target in future searches. 
\end{itemize}

The Legacy Survey of Space and Time \citep{Ivezic2019ApJ...873..111I} at the upcoming Vera C. Rubin Observatory will find many more UFDs and potentially more LMC associated systems \citep[e.g.,][]{Tsiane2025arXiv250416203T}.
Identifying LMC associated UFDs will enable the study of UFDs that formed in a lower density environment.
Stellar spectroscopic follow-up of UFD candidates is key to classify and confirm these systems and to determine their kinematics, orbital properties,  metallicity distributions, and dark matter properties.
The next generation of instruments (e.g., VLT/MOONS\footnote{\url{https://vltmoons.org/}}, VIA\footnote{\url{https://via-project.org/}}, Subaru Prime Focus Spectrograph\footnote{\url{https://pfs.ipmu.jp}}), ongoing and upcoming surveys (e.g., DESI, 4MOST), and the next generation of  30 m-class telescopes will be key to  characterize the distant UFDs that will be found  with Legacy Survey of Space and Time, Euclid, and the Nancy Grace Roman Space Telescope.

\section*{Acknowledgments}

We thank the referee for their comments and feedback.
We thank Andrew Casey and Anna Frebel for co-developing the latest version of SMH. We thank Chris Garling for useful discussions.
APJ was supported by NASA through Hubble Fellowship grant HST-HF2-51393.001 awarded by the Space Telescope Science Institute, which is operated by the Association of Universities for Research in Astronomy, Inc., for NASA, under contract NAS5-26555.
W.C. gratefully acknowledges support from a Gruber Science Fellowship at Yale University. This material is based upon work supported by the National Science Foundation Graduate
Research Fellowship Program under Grant No. DGE2139841. Any opinions, findings, and conclusions or recommendations expressed in this material are those
of the author(s) and do not necessarily reflect the views
of the National Science Foundation.
AC is supported by a Brinson Prize Fellowship at UChicago/KICP.

T.S.Li and G.E.M. acknowledges financial support from Natural Sciences and Engineering Research Council of Canada (NSERC) through grant RGPIN-2022-04794. G.E.M. acknowledges support from an Arts \& Science Postdoctoral Fellowship at the University of Toronto. The Dunlap Institute is funded through an endowment established by the David Dunlap family and the University of Toronto.
The work of V.M.P. is supported by NOIRLab, which is managed by the Association of Universities for Research in Astronomy (AURA) under a cooperative agreement with the U.S. National Science Foundation.
DJS acknowledges support from NSF grant AST-2205863.

This research has made use of the SIMBAD database, operated at CDS, Strasbourg, France \citep{Simbad2000A&AS..143....9W},  NASA's Astrophysics Data System Bibliographic Services, and  the arXiv preprint server.

The DECam Local Volume Exploration Survey (DELVE; NOAO Proposal ID 2019A-0305, PI: Drlica-Wagner) is partially supported by Fermilab LDRD project L2019-011 and the NASA Fermi Guest Investigator Program Cycle 9 No. 91201.
The DELVE project is partially supported by the National Science Foundation under Grant No. AST-2108168 and AST-2307126.

This work has made use of data from the European Space Agency (ESA) mission
{\it Gaia} (\url{https://www.cosmos.esa.int/gaia}), processed by the {\it Gaia}
Data Processing and Analysis Consortium (DPAC,
\url{https://www.cosmos.esa.int/web/gaia/dpac/consortium}). Funding for the DPAC
has been provided by national institutions, in particular the institutions
participating in the {\it Gaia} Multilateral Agreement.

This project used public archival data from the Dark Energy Survey (DES). Funding for the DES Projects has been provided by the U.S. Department of Energy, the U.S. National Science Foundation, the Ministry of Science and Education of Spain, the Science and Technology FacilitiesCouncil of the United Kingdom, the Higher Education Funding Council for England, the National Center for Supercomputing Applications at the University of Illinois at Urbana-Champaign, the Kavli Institute of Cosmological Physics at the University of Chicago, the Center for Cosmology and Astro-Particle Physics at the Ohio State University, the Mitchell Institute for Fundamental Physics and Astronomy at Texas A\&M University, Financiadora de Estudos e Projetos, Funda{\c c}{\~a}o Carlos Chagas Filho de Amparo {\`a} Pesquisa do Estado do Rio de Janeiro, Conselho Nacional de Desenvolvimento Cient{\'i}fico e Tecnol{\'o}gico and the Minist{\'e}rio da Ci{\^e}ncia, Tecnologia e Inova{\c c}{\~a}o, the Deutsche Forschungsgemeinschaft, and the Collaborating Institutions in the Dark Energy Survey.

The Collaborating Institutions are Argonne National Laboratory, the University of California at Santa Cruz, the University of Cambridge, Centro de Investigaciones Energ{\'e}ticas, Medioambientales y Tecnol{\'o}gicas-Madrid, the University of Chicago, University College London, the DES-Brazil Consortium, the University of Edinburgh, the Eidgen{\"o}ssische Technische Hochschule (ETH) Z{\"u}rich,  Fermi National Accelerator Laboratory, the University of Illinois at Urbana-Champaign, the Institut de Ci{\`e}ncies de l'Espai (IEEC/CSIC), the Institut de F{\'i}sica d'Altes Energies, Lawrence Berkeley National Laboratory, the Ludwig-Maximilians Universit{\"a}t M{\"u}nchen and the associated Excellence Cluster Universe, the University of Michigan, the National Optical Astronomy Observatory, the University of Nottingham, The Ohio State University, the OzDES Membership Consortium, the University of Pennsylvania, the University of Portsmouth, SLAC National Accelerator Laboratory, Stanford University, the University of Sussex, and Texas A\&M University.

Based in part on observations at Cerro Tololo Inter-American Observatory, National Optical Astronomy Observatory, which is operated by the Association of Universities for Research in Astronomy (AURA) under a cooperative agreement with the National Science Foundation.

This paper includes data gathered with the 6.5 meter Magellan Telescopes located at Las Campanas Observatory, Chile.

{\it Facilities:} Magellan-Baade (IMACS, \citealt{Dressler2006SPIE.6269E..0FD, Dressler2011PASP..123..288D}), Magellan-Clay (MIKE, \citealt{Bernstein2003SPIE.4841.1694B})

\section*{Software}

\texttt{astropy} \citep{Astropy2013A&A...558A..33A, Astropy2018AJ....156..123A},
\texttt{matplotlib} \citep{matplotlib}, 
\texttt{NumPy} \citep{numpy},
\texttt{iPython} \citep{ipython},
\texttt{SciPy} \citep{2020SciPy-NMeth}
\texttt{corner.py} \citep{corner}, 
\texttt{emcee} \citep{ForemanMackey2013PASP..125..306F},
\texttt{gala} \citep{gala},
\texttt{galpy}\footnote{ http://github.com/jobovy/galpy} \citep{Bovy2015ApJS..216...29B},
\texttt{CarPy} \citep{Kelson2003PASP..115..688K},
\texttt{MOOG} \citep{Sneden1973PhDT.......180S, Sobeck2011AJ....141..175S}
\texttt{SMH} \citep{Casey2014PhDT.......394C}
\texttt{pandas} \citep{reback2020pandas}, 
\texttt{seaborn} \citep{seaborn},
\texttt{pocoMC} \citep{Karamanis2022JOSS....7.4634K, Karamanis2022MNRAS.516.1644K},
{\texttt{NOIRLab IRAF}}\,\citep{Tody1986SPIE..627..733T, Tody1993ASPC...52..173T, Fitzpatrick2024arXiv240101982F}

\section*{Data Availability}

DELVE DR3 photometry will be released in an upcoming data release (Drlica-Wagner in prep.) and the MAGIC photometry with be contained in the first public data release (Chiti et al. in prep). 
Our Magellan/IMACS member catalog is in Table~\ref{table:imacs_members} while the full catalog is in the following zenodo repository: \url{https://doi.org/10.5281/zenodo.15706700}.

\bibliographystyle{aasjournal}
\bibliography{main_bib_file, extra_bib}

\appendix

\section{Search for RR Lyrae stars}
\label{section:rrl}

RR Lyrae (RRL) are variable stars commonly found in old systems such as dwarf galaxies and are excellent distance tracers \citep[e.g.,][]{Beaton2018SSRv..214..113B, Medina2024MNRAS.531.4762M}. RRL have been found in many dwarf spheroidal galaxies and Pic~II with $M_V\sim-2.7$ is predicted to have $N_{\rm RRL}\approx 1$ \citep{MartinezVazquez2019MNRAS.490.2183M, MartinezVazquez2023MmSAI..94d..88M}.
We utilized the {\it Gaia} DR3 RR Lyrae catalog \citep{Gaia_Clementini2023A&A...674A..18C} to search for  RRL in Pic~II. 
There are 3 RRL within 1 degree of Pic~II that are consistent with the proper motion and distance modulus of Pic~II, however, the closest RRL is 33.5\arcmin (or $8.8~R_h$) and all three are more distant than Pic~II (distance modulus $\sim18.4-18.9$ mag).  Due to the large projected distance from the center of Pic~II, we do not consider any of these candidate RRL to be associated with Pic~II. These RRL could be associated with the LMC. 
In addition, the derived photometric metallicities of these three RRL ([Fe/H] $\sim$ -1.7 and -1.9 dex; calculated from \citealt{Mullen2021ApJ...912..144M, Mullen2022ApJ...931..131M} relationships) are more consistent with the LMC or MW halo metallicity than the very metal-poor metallicity of Pic~II. 
The absence of RRL in Pic~II is not surprising as there have been reported a number of faint UFDs without RRL at similar or lower $M_V$  \citep{MartinezVazquez2023MmSAI..94d..88M, Tau2024AJ....167...57T}.

\section{Literature references for figures}
\label{section:appendix_references}

Our references are drawn from the Local Volume Database \citep{Pace2024arXiv241107424P}\footnote{Release 1.0.5 \url{https://github.com/apace7/local_volume_database}}.
References for UFDs including the LMC systems \citep{Bellazzini2005MNRAS.360..185B, Belokurov2007ApJ...654..897B, Bhardwaj2024AJ....167..247B, Boettcher2013AJ....146...94B, Bruce2023ApJ...950..167B, Cantu2021ApJ...916...81C, Carlin2009ApJ...702L...9C, Carlin2017AJ....154..267C, Carlin2018ApJ...865....7C, Casey2025arXiv250104772C, Cerny2021ApJ...910...18C, Cerny2021ApJ...920L..44C, Cerny2023ApJ...942..111C, Cerny2023ApJ...953....1C, Cerny2025ApJ...979..164C, Chiti2021NatAs...5..392C, Chiti2022ApJ...939...41C, Chiti2023AJ....165...55C, Cicuendez2018A&A...609A..53C, Correnti2009MNRAS.397L..26C, Crnojevic2016ApJ...824L..14C, DallOra2006ApJ...653L.109D, DallOra2012ApJ...752...42D, DrlicaWagner2015ApJ...813..109D, Fritz2019A&A...623A.129F, Garofalo2013ApJ...767...62G, Garofalo2025A&A...695A..88G, Greco2008ApJ...675L..73G, Hansen2024ApJ...968...21H, Heiger2024ApJ...961..234H, Homma2018PASJ...70S..18H, Homma2019PASJ...71...94H, Homma2024PASJ...76..733H, Jenkins2021ApJ...920...92J, Ji2021ApJ...921...32J, Karczmarek2015AJ....150...90K, Kim2015ApJ...808L..39K, Kim2016ApJ...833...16K, Kirby2013ApJ...770...16K, Kirby2015ApJ...810...56K, Kirby2017ApJ...838...83K, Koposov2011ApJ...736..146K, Koposov2015ApJ...805..130K, Koposov2015ApJ...811...62K, Koposov2018MNRAS.479.5343K, Kuehn2008ApJ...674L..81K, Lee2009ApJ...703..692L, Li2017ApJ...838....8L, Li2018ApJ...857..145L, Longeard2018MNRAS.480.2609L, MartinezVazquez2015MNRAS.454.1509M, MartinezVazquez2019MNRAS.490.2183M, MartinezVazquez2021AJ....162..253M, MartinezVazquez2021MNRAS.508.1064M, Mateo2008ApJ...675..201M, McConnachie2012AJ....144....4M, Medina2018ApJ...855...43M, Moskowitz2020ApJ...892...27M, Munoz2018ApJ...860...66M, Musella2009ApJ...695L..83M, MutluPakdil2018ApJ...863...25M, MutluPakdil2020ApJ...902..106M, Oakes2022ApJ...929..116O, Pace2020MNRAS.495.3022P, Richstein2022ApJ...933..217R, Richstein2024ApJ...967...72R, Sand2012ApJ...756...79S, Simon2007ApJ...670..313S, Simon2011ApJ...733...46S, Simon2015ApJ...808...95S, Simon2017ApJ...838...11S, Simon2019ARA&A..57..375S, Simon2020ApJ...892..137S, Smith2023AJ....166...76S, Spencer2017ApJ...836..202S, Spencer2018AJ....156..257S, Stetson2014PASP..126..616S, Tan2025ApJ...979..176T, Torrealba2016MNRAS.459.2370T, Torrealba2016MNRAS.463..712T, Torrealba2018MNRAS.475.5085T, Vivas2016AJ....151..118V, Vivas2020ApJS..247...35V, Vivas2022ApJ...926...78V, Walker2009AJ....137.3100W, Walker2015ApJ...808..108W, Walker2015MNRAS.448.2717W, Walsh2008ApJ...688..245W, Wang2019ApJ...881..118W, Willman2006astro.ph..3486W, Willman2011AJ....142..128W}.
References for ambiguous systems \citep{Balbinot2013ApJ...767..101B, Cerny2023ApJ...953....1C, Cerny2023ApJ...953L..21C, Conn2018ApJ...852...68C, Gatto2022ApJ...929L..21G, Homma2019PASJ...71...94H, Kim2015ApJ...799...73K, Kim2015ApJ...803...63K, Kim2016ApJ...820..119K, Luque2018MNRAS.478.2006L, Martin2016ApJ...830L..10M, Mau2019ApJ...875..154M, Mau2020ApJ...890..136M, Munoz2018ApJ...860...66M, Simon2024ApJ...976..256S, Smith2024ApJ...961...92S, Torrealba2019MNRAS.484.2181T}.
References for the GC systems \citep{AlonsoGarcia2025A&A...695A..47A, Bandyopadhyay2025arXiv250504001B, Baumgardt2018MNRAS.478.1520B, Baumgardt2020PASA...37...46B, Baumgardt2021MNRAS.505.5957B, Bradford2011ApJ...743..167B, Caliskan2012A&A...537A..83C, CarballoBello2016MNRAS.462..501C, Crociati2023ApJ...951...17C, DaCosta1992AJ....104..154D, Deras2023ApJ...942..104D, Fanelli2024A&A...688A.154F, Frelijj2025arXiv250218607F, Geisler2023A&A...669A.115G, Gieles2021NatAs...5..957G, GonzalezDiaz2023MNRAS.526.6274G, Hamren2013AJ....146..116H, Hankey2011MNRAS.411.1536H, Harris1996AJ....112.1487H, Henao2025A&A...696A.154H, Johnson2018AJ....155...71J, Joo2019ApJ...875..120J, Jordi2009AJ....137.4586J, Kirby2015ApJ...810...56K, Kobulnicky2005AJ....129..239K, Koch2009A&A...506..729K, Kunder2020AJ....160..241K, Kunder2021AJ....162...86K, Kunder2024AJ....167...21K, Kurtev2008A&A...489..583K, Latour2025A&A...694A.248L, Leanza2024A&A...688A.133L, Longeard2021MNRAS.503.2754L, Loriga2025A&A...695A.156L, Montecinos2021MNRAS.503.4336M, Munoz2018A&A...620A..96M, Munoz2018ApJ...860...66M, Pallanca2023ApJ...950..138P, Richstein2024ApJ...967...72R, Rosenberg1998AJ....115..648R, Saviane2012A&A...540A..27S, Simpson2018MNRAS.477.4565S, Souza2021A&A...656A..78S, Strader2008AJ....136.2102S, Vasquez2018A&A...619A..13V, Voggel2016MNRAS.460.3384V, Weisz2016ApJ...822...32W}.

The chemical abundance references of the UFD stars in Figure~\ref{fig:mike} are as follows: 
Bo\"{o}tes I \citep{Gilmore2013ApJ...763...61G, Frebel2016ApJ...826..110F, Waller2023MNRAS.519.1349W}, 
Bo\"{o}tes II \citep{Ji2016ApJ...817...41J}, 
Canes Venatici II \citep{Francois2016A&A...588A...7F}, 
Carina II and III \citep{Ji2020AJ....160..181J},
Cetus II \citep{Webber2023ApJ...959..141W},
Grus I \citep{Ji2019ApJ...870...83J}, 
Grus II \citep{Hansen2020ApJ...897..183H},
Coma Berenices and Ursa Major II \citep{Frebel2010ApJ...708..560F}, 
Hercules \citep{Koch2008ApJ...688L..13K, Francois2016A&A...588A...7F},
Horologium I \citep{Nagasawa2018ApJ...852...99N}, 
Leo IV \citep{Simon2010ApJ...716..446S,Francois2016A&A...588A...7F}, 
Pisces II \citep{Spite2018A&A...617A..56S}, 
Reticulum II \citep{Ji2016Natur.531..610J, Ji2016ApJ...830...93J}
Segue 1 \citep{Frebel2014ApJ...786...74F}, 
Segue 2 \citep{Roederer2014MNRAS.440.2665R}, 
Triangulum II \citep{Venn2017MNRAS.466.3741V, Kirby2017ApJ...838...83K, Ji2019ApJ...870...83J},
Tucana II \citep{Chiti2018ApJ...857...74C,Chiti2023AJ....165...55C}, 
Tucana III \citep{Hansen2017ApJ...838...44H,Marshall2019ApJ...882..177M},
Tucana V \citep{Hansen2024ApJ...968...21H},
and Ursa Major I \citep{Waller2023MNRAS.519.1349W}.

\end{document}